\documentclass[twocolumn]{aastex7}

\usepackage{xspace}

\newcommand{\emcee}{{\texttt{emcee}}\xspace}
\newcommand{\batman}{{\texttt{batman}}\xspace}
\newcommand{\fleck}{{\texttt{fleck}}\xspace}
\newcommand{\celerite}{{\texttt{celerite}}\xspace}
\newcommand{\ldtk}{{\texttt{PyLDTk}}\xspace}
\newcommand{\BANZAI}{\texttt{BANZAI}\xspace}

\newcommand{\Photutils}{\texttt{Astropy-Photutils}\xspace}
\newcommand{\AstroImageJ}{\texttt{AstroImageJ}\xspace}
\newcommand{\PyPeriod}{\texttt{PyAstronomy-PyPeriod}\xspace}
\newcommand{\lightkurve}{\texttt{Lightkurve}\xspace}

\newcommand{\abc}{Astrobiology Center, 2-21-1 Osawa, Mitaka, Tokyo 181-8588, Japan}
\newcommand{\naoj}{National Astronomical Observatory of Japan, 2-21-1 Osawa, Mitaka, Tokyo 181-8588, Japan}

\newcommand{\sokendai}{Astronomical Science Program, The Graduate University for Advanced Studies (SOKENDAI), 2-21-1 Osawa, Mitaka, Tokyo 181-8588, Japan}

\newcommand{\komaba}{Department of Multi-Disciplinary Sciences, Graduate School of Arts and Sciences, The University of Tokyo, 3-8-1 Komaba, Meguro, Tokyo 153-8902, Japan}
\newcommand{\iac}{Instituto de Astrof\'i sica de Canarias (IAC), 38205 La Laguna, Tenerife, Spain}
\newcommand{\komabasc}{Komaba Institute for Science, The University of Tokyo, 3-8-1 Komaba, Meguro, Tokyo 153-8902, Japan}
\newcommand{\uliege}{Astrobiology Research Unit, Universit\'e de Li\`ege, 19C All\'ee du 6 Ao\^ut, 4000 Li\`ege, Belgium}
\newcommand{\MIT}{Department of Earth, Atmospheric and Planetary Science, Massachusetts Institute of Technology, 77 Massachusetts Avenue, Cambridge, MA 02139, USA}
\newcommand{\harvard}{Center for Astrophysics \textbar \ Harvard \& Smithsonian, 60 Garden Street, Cambridge, MA 02138, USA}

\begin{document}

\title{Multi-band, Multi-epoch Photometry of the Spot-crossing System TOI-3884:\\Refined System Geometry and Spot Properties}

\author[0000-0003-1368-6593, gname=Mayuko, sname=Mori]{Mayuko Mori}
\affiliation{\abc}
\affiliation{\naoj}
\email[show]{mayukomori.519@gmail.com} 

\author[0000-0002-4909-5763, gname=Akihiko, sname=Fukui]{Akihiko Fukui}
\affiliation{\komabasc}
\affiliation{\iac}
\email{afukui@g.ecc.u-tokyo.ac.jp}

\author[0000-0003-3618-7535, gname=Teruyuki, sname=Hirano]{Teruyuki Hirano}
\affiliation{\abc}
\affiliation{\naoj}
\email{hd17156b@gmail.com}

\author[0000-0001-8511-2981,gname=Norio,sname=Narita]{Norio Narita}
\affiliation{\abc}
\affiliation{\komabasc}
\affiliation{\iac}
\email{narita@g.ecc.u-tokyo.ac.jp}

\author[0000-0002-4881-3620,gname=John,sname=Livingston]{John H. Livingston}
\affiliation{\abc}
\affiliation{\naoj}
\affiliation{\sokendai}
\email{john.livingston@nao.ac.jp}


\author[0000-0003-1464-9276,gname=Khalid,sname=Barkaoui]{Khalid Barkaoui}
\affiliation{\iac}
\affiliation{\uliege}
\affiliation{\MIT}
\email{Khalid.Barkaoui@uliege.be}

\author[0000-0001-6588-9574,gname=Karen,sname=Collins]{Karen A.\ Collins}
\affiliation{\harvard}
\email{karen.collins@cfa.harvard.edu}

\author[0000-0002-6424-3410,gname=Jerome,sname='de Leon']{Jerome P. de Leon}
\affiliation{\komabasc}
\email{jpdeleon@g.ecc.u-tokyo.ac.jp}

\author[0000-0002-5978-057X,gname=Kai,sname=Ikuta]{Kai Ikuta}
\affiliation{Graduate School of Social Data Science, Hitotsubashi University, 2-1 Naka, Kunitachi, Tokyo 186-8601, Japan}
\email{kaiikuta.astron@gmail.com} 

\author[0000-0002-0488-6297,gname=Yugo,sname=Kawai]{Yugo Kawai}
\affiliation{\komaba}
\email{yugo6581@g.ecc.u-tokyo.ac.jp}

\author[0000-0001-8227-1020,gname=Richard,sname=Schwarz]{Richard P. Schwarz}
\affiliation{\harvard}
\email{rpschwarz@comcast.net}

\author[0000-0002-1836-3120,gname=Avi,sname=Shporer]{Avi Shporer}
\affiliation{Department of Physics and Kavli Institute for Astrophysics and Space Research, Massachusetts Institute of Technology, Cambridge, MA 02139, USA}
\email{shporer@mit.edu}

\author[gname=Gregor,sname=Srdoc]{Gregor Srdoc}
\affil{Kotizarovci Observatory, Sarsoni 90, 51216 Viskovo, Croatia}
\email{gregorsrdoc@gmail.com}

\begin{abstract}
Spot-crossing transits offer a unique opportunity to probe spot properties such as temperature, size, and surface distribution. TOI-3884 is a rare system in which spot-crossing features are persistently observed during every transit. This is due to its unusual configuration: a {nearly poler orbit} super-Neptune transits a pole-on mid-M dwarf, repeatedly crossing a polar spot. However, previous studies have reported discrepant values in key system parameters, such as stellar inclination and obliquity. To address this, we conducted multi-band, multi-epoch transit observations of TOI-3884b using the MuSCAT instrument series, along with photometric monitoring with the LCO 1m telescopes/Sinistro. We detected time-dependent variations in the spot-crossing signals, indicating that the spot is not exactly on the pole. From the monitoring data, we measured a stellar rotation period of $11.043_{-0.053}^{+0.054}$ days with a modulation amplitude of $\sim$5\% in the $r$-band, consistent with the time variability in the spot-crossing features. Our analysis reconciles previous discrepancies and improves the constraints on the parameters of the system geometry ({$i_\star = 139.9_{-2.0}^{+1.2}$ deg and $\lambda =41.0_{-9.0}^{+3.7}$ deg}) and those of the spot properties (spot radius of $0.425_{-0.011}^{+0.018}\, R_\star$ and spot-photosphere temperature difference of $200_{-9}^{+11}$\,K). These results provide a critical context for interpreting upcoming transmission spectroscopy of TOI-3884\,b, as well as yielding new insights into the magnetic activity and spin-orbit geometry of M dwarfs.
\end{abstract}


\section{Introduction}
With the growing interest in exoplanet characterization, the importance of studying starspots has become increasingly evident. One key reason is that spots can obscure or mimic atmospheric signals in planetary transmission spectra, known as the Transit Light Source (TLS) effect \citep[e.g.,][]{Rackham2018, Rackham2023}. Additionally, spots and stellar activity are known to introduce variability in radial velocity measurements \citep[e.g.,][]{Mahmud2011}, complicating efforts to accurately determine planetary masses.

Despite the significance of these effects, many aspects of spot properties remain uncertain. To estimate their effects, various assumptions about spot temperature, size, and latitude distribution are typically required \citep{Thompson2024}. For M dwarfs in particular, these key spot parameters are not yet well characterized, as empirical measurements remain limited. Of particular interest is the existence of polar spots. Magnetic field studies based on spectroscopic/polarimetric observations have shown that active low-mass stars often exhibit strong dipolar magnetic fields (see reviews by \citealt{Berdyugina2005}, \citealt{Strassmeier2009}, and \citealt{Kochukhov2021}). However, a unified understanding of how such magnetic fields manifest as spots on the photosphere, and what properties those spots exhibit, is still lacking. In particular, while rotational modulation has been widely used to study spot properties \citep[e.g.,][]{Davenport2015}, it is inherently difficult to constrain the spot latitude due to degeneracies in the spot latitude, temperature, and radius.

Another valuable approach to characterizing spots is through spot-crossing events, in which a spot falls within the transit chord and produces a bump in the transit light curve. The size, temperature, and location of a spot can be directly constrained by modeling these light curves \citep[e.g.,][]{Sanchis-Ojeda2011}. Multi-band photometry of spot-crossing events is especially useful for constraining the spot temperature \citep{Schutte2023}.

The TOI-3884 (LSPM J1206+1230, Table~\ref{tab:used_params}) system offers a unique opportunity to characterize the stellar surface via spot-crossing observations. The system consists of a bright M4-dwarf host with a transiting super-Neptune, TOI-3884\,b. This system is particularly unusual because the planet's transits persistently exhibit spot-crossing signals. This is best explained by a {highly misaligned} planet that transits in front of a large spot located near the stellar pole, with the system viewed nearly pole-on. The system was found and confirmed in \citet[][A22 hereafter]{Almenara2022}, and independently characterized by \citet[][L23 hereafter]{Libby-Roberts2023}. 

A22 and L23 both proposed that the spot is located near the stellar pole; however, some of the derived parameter values differed significantly, such as stellar inclination $i_\star$ and projected obliquity (spin-orbit angle) $\lambda$. The major unresolved issue was that neither study succeeded in determining the stellar rotation period, primarily because the light curves taken by the Transiting Exoplanets Survey Satellite \citep[TESS;][]{Ricker2015} did not robustly show rotational brightness modulations. High-resolution spectra provided the projected rotation velocity $v \sin i_\star$ measurements, but the derived $v \sin i_\star$ values from ESPRESSO spectra (1.1km/s, A22) and from HPF spectra ($3.59 \pm 0.92$ km/s, L23) were inconsistent.
Furthermore, since the stellar inclination remained uncertain, it was impossible to derive the rotation period from $v\sin i_\star$ robustly. Strong degeneracies existed between the orbital parameters and the spot parameters, limiting the discussion on the system's architecture and spot properties.

In this paper, we build on these previous results to constrain the spin-orbit geometry of the TOI-3884 system and the properties of its spots with higher precision. We conducted multi-epoch, multi-band transit photometry of TOI-3884b, as well as photometric monitoring of its host star. From the monitoring data, we detect a clear rotational modulation in the stellar brightness. Multiple transit observations also reveal time-variable spot-crossing signals tied to stellar rotation. Additionally, our new multi-band simultaneous spot-crossing measurements allow us to constrain the spot temperature more robustly. As a result, we obtain tighter constraints on both the planetary system's orbital elements and the spot characteristics.

\vskip\baselineskip
This paper is organized as follows: Section~\ref{sec:obs} summarizes the observational data used in our analyses; Section~\ref{sec:rotation} describes our analysis of stellar monitoring light curves; Section~\ref{sec:analyses} describes our analysis of multi-band spot-crossing transit light curves; Section~\ref{sec:discussion} describes the derived TOI-3884 system's spot characteristics and geometry, and compares our results with previous studies. Section~\ref{sec:conclusion} summarizes the results and discussions.

\begin{table*}
\centering
\caption{Parameters of TOI-3884 system which are used in the analyses. \label{tab:used_params}}
\begin{tabular}{llcc}\hline\hline
Name & Units &  Value   & Reference \\ \hline
RA (J2000.0) & hh:mm:ss& 12:06:17.23 & Gaia DR3\\
Dec (J2000.0) & dd:mm:ss & +12:30:25.31& Gaia DR3 \\
Proper motion in RA & mas/yr& $-186.042\pm0.028$& Gaia DR3\\
Proper motion in Dec & mas/yr & $26.388\pm0.017$& Gaia DR3\\
Radial Velocity & km/s & $3.16\pm2.89$& Gaia DR3\\
Parallax & mas & $23.0159\pm0.0260$ & Gaia DR3 \\\\
$G$ magnitude & mag &$14.2462\pm0.0029$ & Gaia DR3 \\
$G_{BP}$ magnitude & mag & $15.9711\pm	0.0047$& Gaia DR3 \\
$G_{RP}$ magnitude & mag &$12.9897\pm	0.0041$& Gaia DR3 \\\\
Stellar radius $R_\star$ & $R_\odot$ & $0.302\pm0.012$ & L23\\
Stellar mass $M_\star$ & $M_\odot$ & $0.298\pm 0.018$ & L23\\
Effective temperature $T_{\rm eff}$ & K & $3180\pm88$ & L23\\ 
Surface gravity $\log g$  & (in cgs) & $4.97\pm0.05$& L23\\
Stellar metallicity  [Fe/H] & dex & $0.04\pm0.12$& L23 \\\hline\hline
\end{tabular}
\end{table*}

\section{Observation and Reduction} \label{sec:obs}

 \subsection{LCO 2m / MuSCAT3, 4}
MuSCAT3 and MuSCAT4 \citep{Narita2020} are multi-band cameras installed on the 2m Faulkes Telescopes operated by Las Cumbres Observatory \citep[LCO;][]{Brown2013} at the Haleakala observatory in Maui, USA, and the Siding Spring observatory in Australia. They have identical optical systems with four channels, enabling simultaneous photometry in the SDSS $g$, $r$, $i$, and $z_s$-bands. Each channel has a 2048$\times$2048 pixel CCD camera with a pixel scale of 0.27\arcsec pix$^{-1}$, providing a field of view of $9'.1 \times 9'.1$. We observed a full transit of TOI-3884\,b with MuSCAT3 on 2024 February 19 and March 31 in UTC, with MuSCAT GTO time. We also observed a full transit with MuSCAT4 on 2024 March 9, under LCO Key Project \citep{Shporer2024}. 

The data reduction was conducted by the standard LCOGT \BANZAI pipeline \citep{McCully2018}. The differential photometry of MuSCAT3 data was conducted using a customized aperture-photometry pipeline for MuSCAT series \citep{Fukui2011}. The optimal aperture radius ($2.7\arcsec$ to $4.3\arcsec$) and a set of comparison stars for each band were selected to minimize the dispersion of the light curves. The differential photometry of MuSCAT4 data was conducted using \AstroImageJ \citep{Collins2017} using optimal aperture radii of $7\arcsec$ to $7.5\arcsec$ for the target and the comparison stars. 

For computational efficiency, we subsequently binned each light curve into 5-min bins. The obtained light curves exhibited different spot-crossing signal shapes, suggesting the possible influence of stellar rotation. This motivated us to conduct monitoring observations of the system.

\subsection{LCO 1m / Sinistro}
Sinistro is a single-band optical camera mounted on worldwide 1m telescopes operated by LCO, with a $26'.5 \times 26'.5$ field of view and a pixel scale of 0.389\arcsec pix$^{-1}$. We observed TOI-3884 from 2024 December 10 to 2025 January 31 and from 2025 March 28 to April 8 with Sinistro in $r$-band with an exposure time of 120 seconds. Two or three frames were taken every 4-8 hours under good observing conditions with airmass $<$ 1.4, moon distance $> 30$ deg, and no weather or instrumental issues. Data reduction was automatically performed with the LCO \BANZAI pipeline.
Photometry was conducted using the package \Photutils \citep{photutils}. We found that circular apertures with a radius of 8 pixels were optimal to reduce the scatter in the light curves. The sky background noise was subtracted using an annulus with an inner radius of 50 pixels and an outer radius of 80 pixels.

We applied differential photometry to correct for telluric variability. In principle, the telluric effect on stellar brightness variations should depend on the stellar color, given that the degree of atmospheric extinction depends on the wavelength. 
To mitigate this effect, we applied a color-magnitude correction. Specifically, we obtained the $G_{BP}-G_{RP}$ colors of stars in the field of view from the Gaia DR3 catalog \citep{Gaia2016, Gaia2023}. For each frame, we fitted a linear function to the relationship between the stellar colors and the relative measured magnitudes with respect to a reference frame, using all stars that were matched with the catalog within the field of view. We then corrected the relative magnitude of the target star, TOI-3884, by subtracting the fitted value corresponding to its color ($G_{BP}-G_{RP}=2.98$).

We excluded frames with the following three conditions as outliers: (i) the number of detected stars was smaller than 50; (ii) the median stellar magnitude was $> 0.5$ mag fainter than that of the reference frame; (iii) the obtained target brightness was more than $3\sigma$ different from the mean model obtained by fitting a third-order polynomial. The light curve was then binned into 0.5-day intervals. 

\begin{figure*}
    \centering
\includegraphics[width=1\linewidth]{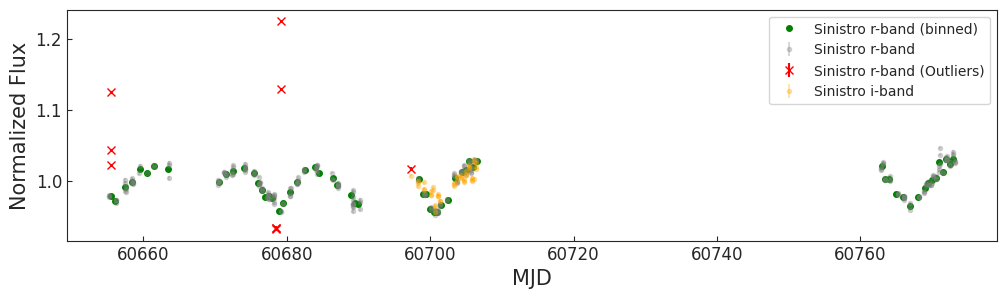}
    \caption{Monitoring light curve of TOI-3884 obtained with Sinistro. The horizontal axis indicates the Modified Julian Date (MJD). Gray points represent the $r$-band photometric measurements, green points show the binned light curve in 0.5-day intervals, and red crosses mark the data points that were removed as outliers. The yellow points indicate the $i$-band measurements.}
    \label{fig:monitoring}
\end{figure*}

Figure~\ref{fig:monitoring} shows the obtained light curve, which exhibits periodic brightness modulation. The outliers are marked by red crosses. These include two brightening events at MJD$\sim$60655 and 60679, which appeared to be flares. A low flux anomaly at MJD$\sim$60678.5 was due to the transit of TOI-3884\,b. 

Note that we also observed the system with Sinistro in $i$-band, from 2025 January 20 to 31, with 20-sec exposure times. Following the same photometry procedure as for the $r$-band data, the $i$-band light curve also exhibited a clear modulation feature with a smaller amplitude (Figure~\ref{fig:monitoring}). This strengthens the evidence that the observed variations originate from an astronomical source rather than systematics. However, due to the short duration of the $i$-band observations, we did not use this dataset in subsequent analyses.

\section{Monitoring light curve Analyses} \label{sec:rotation}

The sinusoidal-like variations observed in the Sinistro $r$-band light curve are likely caused by the rotation of a spotted star. Even if the spot remains constantly visible, stellar rotation can still produce brightness modulations due to changes in the projected spot shape and the effect of limb darkening if the spot is not located exactly at the stellar pole. We investigated the stellar rotation period using two different methods: the Generalized Lomb-Scargle periodogram \citep[GLS;][]{GLS} and Gaussian process (GP) modeling \citep[e.g.,][]{Foreman-Mackey2017F, Angus2018}.

\begin{figure*}
    \centering
\includegraphics[width=1\linewidth]{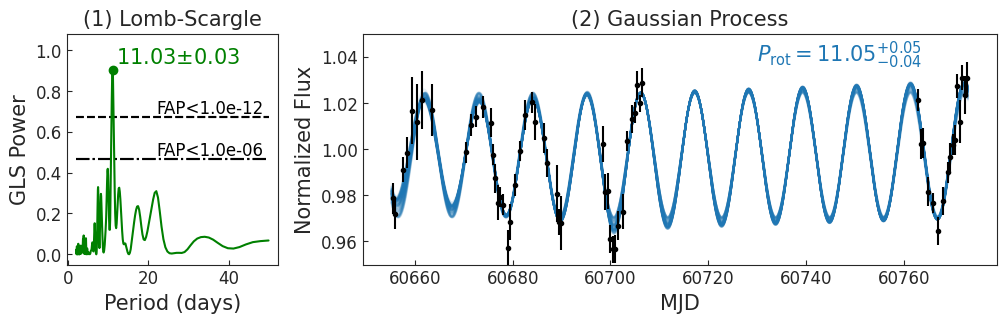}
    \caption{{\it left.} GLS periodogram of the monitoring light curve observed in Sinistro $r$-band. A strong peak is detected at $11.03\pm0.03$ days. The dashed horizontal lines indicate the false alarm probability (FAP) corresponding to the peak height.\\{\it right.} GP models (blue lines), generated using 100 randomly selected parameter sets from the converged \texttt{emcee} samples, overlaid on the observed light curve (black points). The error bars are scaled by the best-fit $\alpha$ factor.}
    \label{fig:GLSGP}
\end{figure*}

First, we analyzed the periodic signals in the Sinistro $r$-band light curve using GLS as implemented in the Python package \texttt{PyPeriod} \citep{pya}. We searched for periodicities in the range of 2 to 50 days and identified a strong GLS power peak at $11.03\pm0.03$ days (Figure~\ref{fig:GLSGP}). The false alarm probability (FAP) was below $10^{-12}$, indicating a robust detection.

To verify this result independently and estimate the uncertainty in the rotation period based on Bayesian sampling, we also applied a Gaussian process model. Quasi-periodic kernels have been widely used and demonstrated to be robust for inferring stellar rotation periods from photometric time series \citep[e.g.,][]{Foreman-Mackey2017F, Angus2018}. We adopted the modified quasi-periodic kernel implemented in \texttt{celerite} (Equation 56 in \citealt{Foreman-Mackey2017F}),
\begin{equation}
    k(\tau) = \frac{B}{2+C}e^{-\tau/L}\Big[\cos\left(\frac{2\pi\tau}{P_{\rm rot}}\right)+(1+C)\Big],
\end{equation}
where $B$ controls the overall amplitude of variations, 
$C$ determines the relative strength of the periodic component, $L$ sets the decay timescale of correlations, and $P_{\rm rot}$ defines the rotation period. 
In addition to these four hyperparameters, we introduced a scaling factor $\alpha$ to account for white noise. Specifically, the photometric uncertainties were scaled as $\alpha \times err$, where $err$ represents the original measurement uncertainties. We performed parameter sampling using Markov Chain Monte Carlo (MCMC) with \texttt{emcee} \citep{Foreman-Mackey2013}. The sampling was conducted for the parameters $\alpha$, $\log B$, $\log C$, $\log L$, and $\log P_{\rm rot}$, each given a broad uniform prior.
The derived parameter values and their uncertainties are based on the 16th, 50th, and 84th percentiles of the posterior distributions. They are summarized in Table~\ref{tab:GP}, and Figure~\ref{fig:GLSGP} presents the resulting best-fit GP model overlaid on the observed light curve.
The derived rotation period was $P_{\rm rot}=11.05_{-0.04}^{+0.05}$ days, which is consistent with the GLS result. We adopt this value as a prior in the following transit analysis.

\begin{table}
\caption{Best-fit values and uncertainties for the GP fit parameters, derived from the posterior distributions. \label{tab:GP}}
\centering
\begin{tabular}{lcc}\hline\hline
Parameter & Prior$^\ast$ & Derived Value \\ \hline
$\alpha$& $\mathcal{U}$(0, 20) & $7.5_{-0.9}^{+1.0}$\\
$\log B$& $\mathcal{U}$($-10, 10$) & $-7.3_{-0.8}^{+1.1}$\\
$\log C$& $\mathcal{U}$(-10, 10)& $-5.0 _{-3.4}^{+3.8}$\\
$\log L$& $\mathcal{U}$(-20, 2)& $-7.6_{-2.8}^{+1.4}$\\
$\log P_{\rm rot}$ & $\mathcal{U}$(0, 5) & $2.4020_{-0.0038}^{+0.0049}$\\\hline\hline
\multicolumn{3}{l}{$^\ast$ $\mathcal{U}$(A, B) indicates the uniform prior from A to B.} 
\end{tabular}
\end{table}

\section{Transit light curve Analyses} \label{sec:analyses}

We simultaneously modeled the 12 transit light curves of TOI-3884\,b, observed with MuSCAT3 and MuSCAT4 in four bands during three separate observations. We used a spot-crossing transit model multiplied by a baseline model, which was represented by a third-order polynomial in time.

For the spot-crossing models, we used the package \fleck \citep{Morris2020}, which utilizes the transit model \batman \citep{Kreidberg2015}. \fleck defines a spot by its contrast $c_B$, radius $r_{\rm spot}$, latitude $\theta_{\rm spot}$, and longitude $\varphi_{\rm spot}$, assuming a disk-like spot placed on the stellar surface, and computes its contribution to the total stellar flux. We assumed that the spot parameters remained constant across the three observed transits, i.e., the spot configuration did not change over the $\sim$40-day span between the transit observations, and only its phase angle changed due to stellar rotation. This assumption is supported by the stable shape of the rotational modulation observed over $\sim$100 days, as in Figure~\ref{fig:monitoring}. We also fixed the number of spots to one, as the MCMC chains failed to converge when multiple spots were introduced into the model.

We treated the photosphere temperature $T_{\rm phot}$ and the temperature difference between the photosphere and the spot, defined as $\Delta T \equiv T_{\rm phot} - T_{\rm spot} $, as free parameters, and calculated the corresponding spot contrast $c_B$ in each bandpass for each temperature combination using our custom module, \texttt{spocon}\footnote{\texttt{spocon}: \url{https://github.com/Mayuko-Mori/spocon}}.
{In \texttt{spocon}, the spot contrast $c_B$, defined as the relative flux contribution of spots to the photosphere in each bandpass, is calculated as:
\begin{equation} 
c_B = \frac{\int_{\rm B} d\lambda \mathcal{T}_\lambda \mathcal{F}_{\lambda, \rm{spot}}} {\int_{\rm B} d\lambda \mathcal{T}_\lambda \mathcal{F}_{\lambda, \rm{phot}}}, \label{eq:morris}
\end{equation}
where $\mathcal{T}_\lambda$ is the filter transmission function and $\mathcal{F}_{\lambda, \rm{spot/phot}}$ are the stellar spectra from the spots or the photosphere at the temperature of $T_{\rm spot}$ and $T_{\rm phot}$, respectively.} As the stellar spectra, \texttt{spocon} employs the BT-Settl stellar spectral models\footnote{Theoretical Model Services: \url{http://svo2.cab.inta-csic.es/theory/main/}} \citep{Allard2014}. We used stellar models with surface gravity $\log g=5$ in cgs and metallicity [Fe/H]$= 0$ dex.

For the transit parameters, we adopt the transit-center time $T_0$, the orbital period $P_{\rm orb}$, the orbital semi-major axis over the stellar radius $a/R_\star$, the impact parameter $b$, and the ratio of the planetary radius to the stellar radius $k = R_p/R_\star$. The parameter $k$ was independently set for each band. To account for the change in the spot's position between transits, we introduced the stellar rotation period $P_{\rm rot}$, the stellar inclination $i_\star$, and the projected obliquity between the stellar-spin axis and orbital axis $\lambda$ as parameters. In addition, modified quadratic limb-darkening coefficients $(q_1, q_2)$ were set as parameters for each band \citep{Kipping2013}. The total number of parameters was 24. Note here that we assumed a circular orbit, as suggested by previous works. For completeness, we also tested transit modeling with eccentricity and argument of periapsis as free parameters, and found that $e$ was consistent with zero and that the posterior distributions of other parameters remained unchanged, confirming that fixing $e=0$ is justified.

We applied Gaussian priors to $T_{\rm phot}$, $T_0$, $P_{\rm orb}$, and $R_p/ R_\star$, using the values derived from L23. Instead of $a/R_\star$ and $b$, we calculated the stellar density $\rho_\star$ and transit duration $T_{14}$ and applied Gaussian priors for them based on the values from L23, as they allow for more physically reasonable priors. We also applied a Gaussian prior to $P_{\rm rot}$ from our monitoring observations result.

For the limb-darkening coefficients, we assigned Gaussian priors based on theoretical calculations using the \ldtk package \citep{Parviainen2015, Husser2013}, assuming a stellar effective temperature of $ T_{\rm eff} = 3180 \pm 88 \,{\rm K} $, metallicity $ {\rm [Fe/H]} = 0.04 \pm 0.12 \,{\rm dex} $, and stellar surface gravity $ \log g = 4.97 \pm 0.05 $ (cgs). To avoid overly constraining the limb-darkening coefficients, we set a large standard deviation ($\sigma (q_1)=0.1$, $\sigma (q_2)=0.05$) for the Gaussian priors. Uniform priors were applied to the remaining parameters, or used in combination with Gaussian priors where appropriate, with bounds chosen to avoid unphysical values. 

To correct for the potential under/over estimation of the flux uncertainties, we first conducted a $\chi^2$-minimization fitting and corrected the flux uncertainties so that the reduced $\chi^2$ became unity. The posterior sampling was performed using \emcee, adopting the differential evolution algorithm with a snooker updater \citep{Braak2008}, which is suited for exploring multi-modal posterior distributions.

{From the MCMC results, we found that the observed spot-crossing light curves can be explained by several distinct solutions. In particular, the stellar inclination $i_\star$ clusters around two regions: approximately $-40$\,deg and $140$\,deg. These correspond to whether the stellar rotation is counterclockwise or clockwise as seen from the observer's perspective—that is, whether the visible pole is the north or south pole.

The solution with $i_\star \sim -40$\,deg is associated with a stellar rotation period of approximately 11.25 days. In contrast, the solution with $i_\star \sim 140$\,deg further splits into two groups: one with $P_{\rm rot} \sim 11.04$ days and the other with $P_{\rm rot} \sim 11.5$ days. The solutions near 11.25 and 11.5 days exhibited log-probability values more than 30 and 50 lower, respectively, than those near 11.04 days, indicating significantly worse fits.

To eliminate these less probable configurations, we imposed a uniform prior of $\mathcal{U}(90, 180)$ on $i_\star$, and $\mathcal{U}(10.8, 11.3)$ on $P_{\rm rot}$, the latter restricting the range to within $5\sigma$ of the value inferred from the rotational modulation. After applying these priors and rerunning the MCMC, the chains successfully converged.}

\begin{figure*}
    \centering
\includegraphics[width=1\linewidth]{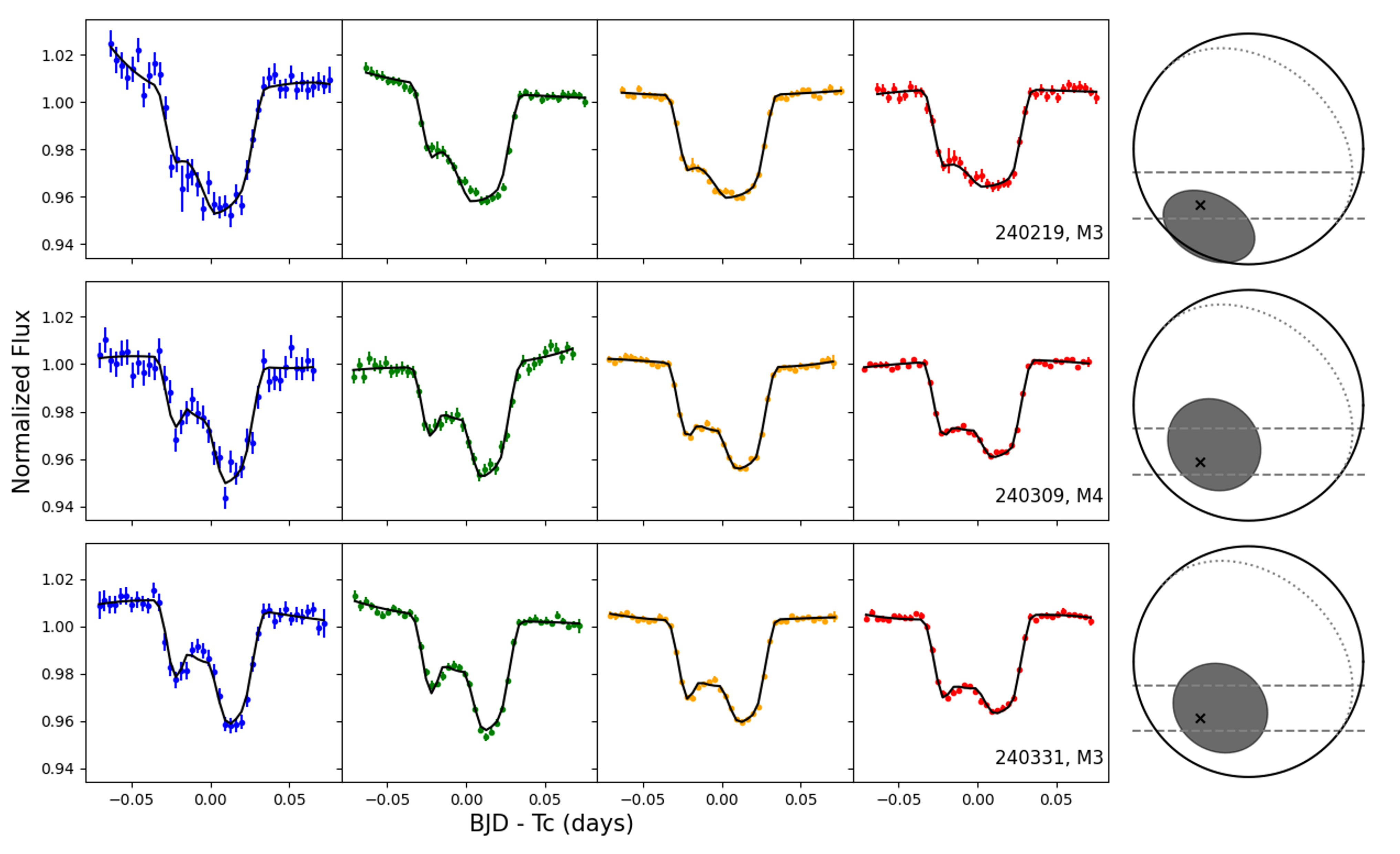}
    \caption{Observed transit light curves of TOI-3884b (top to bottom: three transits observed with MuSCAT3 or MuSCAT4; left to right: $g$-, $r$-, $i$-, and $z$-band) and the corresponding best-fit models (black lines). The error bars in the datapoints are scaled to make reduced $\chi^2$ values equal to 1. To the right of each light curve, the spot distribution on the stellar surface at the corresponding epoch is shown, as modeled with \fleck. The x mark in the figure shows the location of the stellar {south} pole, the dotted lines show the stellar equator, and the regions inside dashed lines indicate the transit chord.}
    \label{fig:transit_best}
\end{figure*}

Even after convergence, the posterior distribution appeared to contain two slightly different solutions with respect to parameters such as $\lambda$. However, these variations are within the 2$\sigma$ uncertainty and primarily contribute to broadening the uncertainties of the parameters. In Appendix~\ref{ap:comp}, we further investigate these degenerate solutions and present the posterior distributions of each parameter. The priors and derived parameter values are summarized in Table~\ref{tab:retrieved_params}, and the best-fit spot-crossing transit model is shown in Figure~\ref{fig:transit_best}.

\begin{table*}
\centering
\caption{Parameters of TOI-3884 system. \label{tab:retrieved_params}}
\begin{tabular}{llcc}\hline\hline
Name & Parameter [Units] & Prior$^\ast$ & Derived Value \\
\hline
\multicolumn{4}{l}{\it $\diamond$ Sampled parameters} \\
Photosphere temperature & $T_{\rm phot}$ [K] & $\mathcal{U}$(2700, 5000) $\times$ $\mathcal{N}$(3180, 88) & $3151_{-42}^{+45}$\\
Spot temperature difference & $\Delta T$ [K] & $\mathcal{U}$(0, 1000) & $200_{-9}^{+11}$\\
Spot radius & $r_{\rm spot} [R_\star]$ & $\mathcal{U}$(0, 0.5)& $0.425_{-0.011}^{+0.018}$\\
Spot latitude & $\theta_{\rm spot}$ [deg] & $\mathcal{U}$(-90, 90)& {$-76.8_{-4.6}^{+2.5}$}\\
Spot longitude & $\varphi_{\rm spot}$ [deg] & $\mathcal{U}$(0, 360)& {$262_{-14}^{+6}$}\\\\
Transit center & $T_0 - 2460360$ [BJD]& $\mathcal{N}$(0.90785, 0.00025) & $0.908017 \pm 0.000078$\\
Orbital period & $P_{\rm orb}$ [days] & $\mathcal{N}$(4.5445828, 0.0000098)& $4.5445858_{-0.0000086}^{+0.0000085}$ \\
Semimajor axis & $a/R_\star$ & $\mathcal{U}$(0, 100) & $23.47 \pm 0.20$\\
Impact parameter & $b$ & $\mathcal{U}$(0, 1) & $0.402_{-0.019}^{+0.018}$ \\
Planetary radius & $k = R_p/R_\star$~($g$)& $\mathcal{U}$(0, 1) $\times$ $\mathcal{N}$(0.197, 0.002) & $0.1978 \pm 0.0018$\\
& $R_p/R_\star$~($r$)& $\mathcal{U}$(0, 1) $\times$ $\mathcal{N}$(0.197, 0.002) & $0.1957\pm0.0015$\\
& $R_p/R_\star$~($i$)& $\mathcal{U}$(0, 1) $\times$ $\mathcal{N}$(0.197, 0.002) & $0.1924\pm0.0012$\\
& $R_p/R_\star$~($z$)& $\mathcal{U}$(0, 1) $\times$ $\mathcal{N}$(0.197, 0.002) & $0.1908\pm0.0012$\\\\
Rotation period & $P_{\rm rot}$ [days] & $\mathcal{U}$(10.8, 11.3) $\times$ $\mathcal{N}$(11.05, 0.05)& $11.043_{-0.053}^{+0.054}$ \\
Stellar inclination & $i_\star$ [deg] & {$\mathcal{U}$(90, 180)}& {$139.9_{-2.0}^{+1.2}$}\\
Projected obliquity & $\lambda$ [deg] & $\mathcal{U}$(-180, 180)& {$41.0_{-9.0}^{+3.7}$}\\
limb-darkening & $q_1$~($g$) & $\mathcal{U}$(0, 1) $\times$ $\mathcal{N}$(0.78, 0.10) &  $0.509_{-0.079}^{+0.080}$\\
coefficients & $q_2$~($g$) &  $\mathcal{U}$(0, 1) $\times$ $\mathcal{N}$(0.33, 0.05) &  $0.303 \pm 0.048$\\
& $q_1$~($r$) & $\mathcal{U}$(0, 1) $\times$ $\mathcal{N}$(0.67, 0.10) & $0.284_{-0.056}^{+0.059}$ \\
& $q_2$~($r$) &  $\mathcal{U}$(0, 1) $\times$ $\mathcal{N}$(0.36, 0.05) & $0.323 \pm 0.049$ \\
& $q_1$~($i$) &  $\mathcal{U}$(0, 1) $\times$ $\mathcal{N}$(0.47, 0.10) & $0.215_{-0.044}^{+0.047}$ \\
& $q_2$~($i$) &  $\mathcal{U}$(0, 1) $\times$ $\mathcal{N}$(0.28, 0.05) & $0.272\pm0.048$\\
& $q_1$~($z$) &  $\mathcal{U}$(0, 1) $\times$ $\mathcal{N}$(0.39, 0.10) &  $0.153_{-0.035}^{+0.039}$\\
& $q_2$~($z$) &  $\mathcal{U}$(0, 1) $\times$ $\mathcal{N}$(0.23, 0.05) &  $0.286\pm0.048$\\
\hline
\multicolumn{4}{l}{\it $\diamond$ Parameters used as conditioning priors} \\
Stellar density & $\rho_\star$ [g/cm$^3$] & $\mathcal{N}$(15.26, 2.04) &  $11.85_{-0.29}^{+0.30}$  \\
Transit duration & $T_{14}$ [days] & $\mathcal{N}$(0.0666,0.0024) & $0.06926\pm0.00031$\\\hline
\multicolumn{4}{l}{\it $\diamond$ Derived parameters} \\
Spot temperature & $T_{\rm spot}$ [K] & - & $2952\pm36$ \\
Planetary radius$^a$ & $R_p$ [$R_\oplus$] & -  & $6.370\pm0.022$\\
Orbital inclination & $i_{\rm orb}$  [deg] & - & $89.018_{-0.051}^{+0.053}$\\
True obliquity & $\Psi$ [deg] & - & {$61.9_{-5.6}^{+2.3}$}\\
Projected rotation velocity & $v \sin i_\star$ [km/s] & - & $0.892_{-0.024}^{+0.038}$ \\
\hline\hline
\multicolumn{4}{l}{$^\ast$ $\mathcal{U}$(A, B) indicates the uniform prior from A to B, and $\mathcal{N}$(X, Y) indicates the Gaussian prior with the mean value X}\\ \multicolumn{4}{l}{~~and standard deviation Y.}\\
\multicolumn{4}{l}{$^a$ Calculated from weighted mean of $R_p$ in the four bands.}
\end{tabular}
\end{table*}

\section{Discussion} \label{sec:discussion}

Table~\ref{tab:comparison} summarizes a comparison of the system parameters derived from our observations with those reported in two previous studies, A22 and L23. In the following sections, we examine the stellar rotation, spot characteristics, and the system’s spin-orbit geometry, incorporating comparisons with the previous works.

\begin{table*}
\centering
\caption{Parameter value comparison with A22 \citep{Almenara2022} and L23\citep{Libby-Roberts2023}.}\label{tab:comparison}
\begin{tabular}{llccc}\hline\hline
Name & Parameter [Units] & A22 & L23 & {\bf This work} \\
\hline
Photosphere temperature & $T_{\rm phot}$ [K] &  $3249\pm74$$^a$ & $3180 \pm 88$  & $3151_{-42}^{+45}$\\
Spot temperature difference$^b$ & $\Delta T$ [K] & $187 \pm 21$  & 200-500$^c$ & $200_{-9}^{+11}$\\
Spot temperature$^b$ & $T_{\rm spot}$ [K] &  $3062\pm77$$^c$  &  2700-2900  & $2952\pm36$ \\
Spot radius$^b$ & $r_{\rm spot} [R_\star]$ &  $0.75\pm0.05$$^c$ &  0.29, 0.16, 0.09$^d$  & $0.425_{-0.011}^{+0.018}$\\
Spot latitude$^b$ & $\theta_{\rm spot}$ [deg]  & 90 (fixed) &   -   &  {$-76.8_{-4.6}^{+2.5}$}\\
Spot longitude$^b$ & $\varphi_{\rm spot}$ [deg]  & -  &   -   &  {$262_{-14}^{+6}$}\\\\
Transit center & $T_0-2 450 000$ [BJD] & $9642.86314 \pm 0.00012$ & $9556.51669\pm0.00025$ &  $10360.908017 \pm 0.000078$\\
Orbital period & $P_{\rm orb}$ [days] & $4.5445697 \pm 0.0000094$  &  $4.5445828 \pm 0.0000098$ & $4.5445858_{-0.0000086}^{+0.0000085}$ \\
Stellar density & $\rho_\star$ [g/cm$^3$] & $14.3 \pm 1.1$  &  $15.26 \pm 2.04$ &  $11.85_{-0.29}^{+0.30}$  \\
Planetary radius & $R_p$ [$R_\oplus$] &  $6.00 \pm 0.18$ &  $6.43 \pm 0.20$ & $6.370\pm0.022$\\
Semimajor axis & $a/R_\star$  &  $25.01 \pm 0.65$ &  $25.90^{+0.96}_{-0.71}$ &   $23.47 \pm 0.20$\\
Impact parameter & $b$ & $-0.04 \pm 0.13$  & $0.089^{+0.082}_{-0.061}$  &   $0.402_{-0.019}^{+0.018}$ \\
Orbital inclination & $i_{\rm orb}$  [deg] & $90.10 \pm 0.29$  &  $89.81^{+0.13}_{-0.18}$ & $89.018_{-0.051}^{+0.053}$\\\\
Rotation period & $P_{\rm rot}$ [days] &  -  &  -  & $11.043_{-0.053}^{+0.054}$ \\
Projected rotation velocity & $v \sin i_\star$ [km/s] &  $1.1$ & $3.59 \pm 0.92$  & $0.892_{-0.024}^{+0.038}$ \\\\
Stellar inclination & $i_\star$ [deg] &  $47.0 \pm 8.5$   &  $25\pm 5$   & {$139.9_{-2.0}^{+1.2}$}\\
Projected obliquity & $\lambda$ [deg] &  $151 \pm 11$   &  $75\pm10$   & {$41.0_{-9.0}^{+3.7}$}\\
True obliquity & $\Psi$ [deg] & {$50 \pm12$ or $130\pm12$} &   $84\pm5$$^c$ & {$61.9_{-5.6}^{+2.3}$}\\
\hline\hline
\multicolumn{5}{l}{$^a$ From there spot contrast analysis (Appendix B in A22).}\\
\multicolumn{5}{l}{$^b$ Spot parameters could vary significantly depending on the observation timing, considering possible spot evolution.}\\
\multicolumn{5}{l}{$^c$ To ensure consistency in formatting, we calculated the values based on those reported in their paper.}\\
\multicolumn{5}{l}{$^d$ L23 assumed three spot model.}\\
\end{tabular}
\end{table*}

\subsection{Stellar rotation}\label{sec:dis-rotation}

From the monitoring observations, we successfully detected the rotational brightness variations of TOI-3884 for the first time and obtained a rotational period of $\sim 11$ days. This period also explains the shape variations of the spot-crossing signals observed in transit light curves. Additionally, we examined archival data from the Zwicky Transient Facility (ZTF) covering over six years of monitoring of TOI-3884. While the signal was not strong, we also found hints of an $\sim$11-day modulation in certain periods (see Appendix~\ref{ap:ZTF} for details). This result supports the robustness of our rotational period detection.

\begin{figure}
    \centering
\includegraphics[width=1\linewidth]{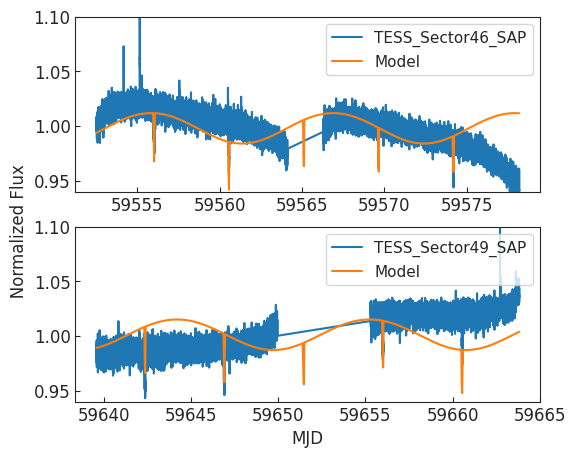}
    \caption{SAP light curves of TOI-3884 from TESS Sectors 46 and 49, overlaid with the forward model (orange lines) of the rotational modulation generated using the best-fit parameters from the transit fitting with MuSCAT3/4 data. The discrepancy between the model and the TESS light curves suggests that the spot configuration may have been different during the TESS observations compared to the MuSCAT3/4 observations.}
    \label{fig:TESS_monitoring}
\end{figure}

However, this modulation was not observed in the TESS light curves, as previously reported in A22 and L23. To further investigate this, we reanalyzed the TESS light curves obtained in Sector 46 and Sector 49, and processed by the TESS Science Processing Operations Center \citep[SPOC;][]{Jenkins2016}. A period search using the GLS method on the TESS Simple Aperture Photometry (SAP) light curves did not detect the 11-day period. Given that we observed $\sim$5\% variability in the $r$-band, assuming a spot contrast of $\Delta T \sim 200$ K, we would expect $\sim$3\% variability in the TESS light curve. However, no such variation was detected (Figure~\ref{fig:TESS_monitoring}).

There are two possible explanations for this discrepancy. The first possibility is that systematic trends in the TESS light curves masked the rotational modulation. Indeed, the SAP light curves show a systematic trend with a period of 12–14 days. While this alone is unlikely to completely obscure the rotational modulation, it may have made it more difficult to detect small-amplitude variations. The second possibility is that the spot characteristics changed between the TESS observations and our Sinistro observations. Given that there is a three-year gap between these observations, it is reasonable to expect variations in the spot distribution. The presence of spot-crossing signals in the TESS transit light curves {which were analyzed in A22 and L23,} suggests that spots were still present during the TESS observations. However, if the spots were located closer to the stellar pole, the rotational modulation could have been reduced, which is a plausible scenario. In fact, we found that the TESS transit light curves showed little variation in the spot-crossing signals and that the data can be explained by an exact-polar spot (Appendix~\ref{ap:TESS}). The ZTF light curves also suggest that the amplitude of brightness variations was smaller during the TESS observation period (see Appendix~\ref{ap:ZTF}). To confirm the timescale over which the spot distribution on TOI-3884 changes, long-term monitoring and contemporaneous transit observations are essential.

In addition, based on the stellar rotation period obtained in this study, along with the stellar radius and inclination, we calculate $v \sin i_\star$ to be $0.892_{-0.024}^{+0.038}$~km/s. 
Our result is closer to that of A22, which derived a value of 1.1 km/s. Given that rotational broadening below $\sim$1 km/s is close to the detection limit for both ESPRESSO \citep{Pepe2021} and HPF \citep{Mahadevan2012}, we interpret the previously reported $v \sin i_\star$ values as upper limits.

\subsection{Connection between stellar rotation and polar spot}

This work has revealed the $\sim$11-day rotation period of TOI-3884 and confirmed the presence of a large spot located near, but not exactly at, the stellar pole. This updated picture of the system enables more detailed investigations of the star’s magnetic activity, including comparisons with theoretical predictions.

Polar spots have been commonly observed in rapidly rotating stars, particularly through Doppler imaging techniques \citep{Berdyugina2005}. Both theoretical studies and polarization observations suggest that M dwarfs with deep convective zones tend to exhibit dominant bipolar (poloidal) magnetic fields when they rotate rapidly \citep{Yadav2015, Moutou2017}. In light of this, we investigated whether the large polar spot on TOI-3884 can be explained by its rotation.

The Rossby number, defined as $Ro = P_{\rm rot} / \tau_{\rm conv}$, is a key parameter characterizing how strongly stellar magnetic activity is influenced by rotation, where $\tau_{\rm conv}$ is the convective turnover timescale. Based on the empirical relation between stellar mass and $\tau_{\rm conv}$ (Equation 6 of \citealt{Wright2018}), {and adopting a stellar mass of $0.298 \pm 0.018M_\odot$ derived from high-resolution spectroscopy (L23)}, we estimate $\tau_{\rm conv}$ for TOI-3884 to be {60–110 days. Combined with the measured stellar rotation period of $\sim 11$\,days, this yields a Rossby number of $Ro = 0.10$–$0.18$}. According to \citet{Yadav2015}, polar spot formation is favored when $Ro \ll 1$. Although TOI-3884's Rossby number is somewhat larger than those considered in their model, it still lies within the rotation-dominated regime. Independent measurements of the magnetic field strength in TOI-3884 would provide further insight into this scenario.

\subsection{Spot Temperature}

From our simultaneous multi-band spot-crossing transit light curve analysis, we measured a photospheric temperature of $T_{\rm phot}=3151_{-42}^{+45}$\,K and a photsphere-spot temperature difference of $\Delta T= 200_{-9}^{+11}$\,K, which corresponds to a spot temperature of $T_{\rm spot}=2952\pm36$\,K.

The resulting spot temperature is consistent with the value derived by A22, who estimated the spot temperature from multi-band spot-crossing events. In their study, however, the data were obtained with different instruments and at various epochs, which could introduce additional uncertainties due to potential spot evolution. Our observations enable a more robust and self-consistent measurement of the spot temperature.

\begin{figure}
    \centering
\includegraphics[width=1\linewidth]{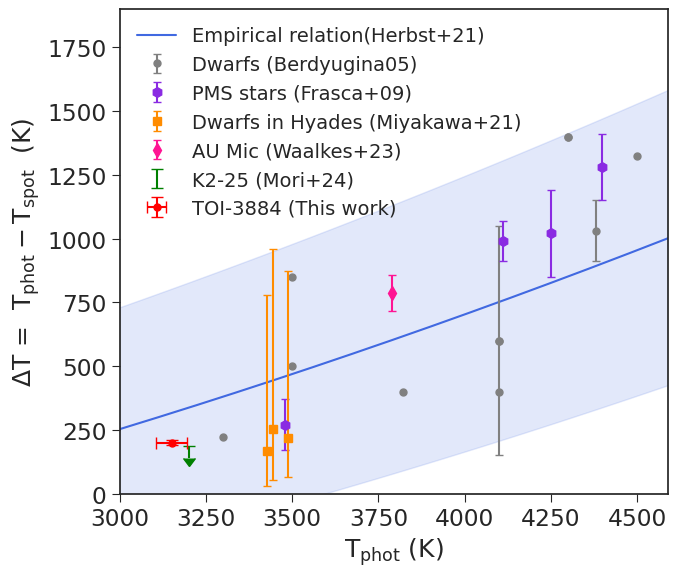}
    \caption{{$T_{\rm phot}$-$\Delta T$ relation for dwarf stars with $T_{\rm phot}$ ranging from 3000\,K to 4500\,K. The empirical relation and its uncertainties from Equation~4 in \citet{Herbst2021} (originally derived by \citealt{Berdyugina2005}) are shown as the blue line and shaded region. Error bars for $T_{\rm phot}$ are omitted for all stars except TOI-3884. For data points without error bars in $\Delta T$, uncertainties were not reported.}}
    \label{fig:spot_temp}
\end{figure}

{Figure~\ref{fig:spot_temp} compares the $T_{\rm phot}$ and $\Delta T$ values derived for TOI-3884 with the empirical relation and with previous measurements from other stars. Our result agrees with the empirical relation proposed by \citet{Herbst2021} (originally derived by \citealt{Berdyugina2005}). At the time this relation was derived, spot temperature measurements in M dwarfs were limited, resulting in large uncertainties. Although such measurements remain scarce, our result adds to the small but growing number of well-characterized spot temperatures in M dwarfs \citep[e.g.,][]{Frasca2009, Miyakawa2021, Waalkes2024, Mori2024}, and will contribute to refining the empirical relation in future studies.}

\subsection{Transit Depth Chromaticity}

\begin{figure}
    \centering
\includegraphics[width=0.8\linewidth]{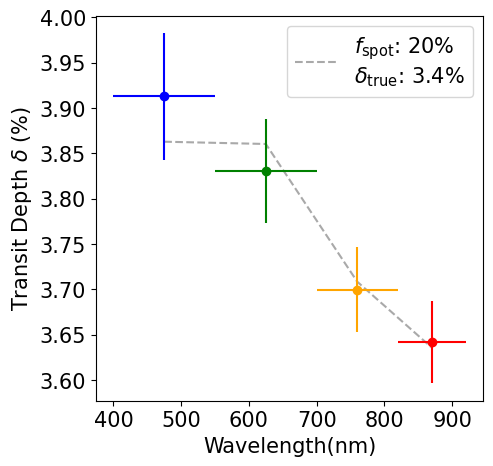}
    \caption{Derived transit depth of TOI-3884 in each band ($g$, $r$, $i$, and $z$-band from left to right). The dashed line indicates the expected chromatic variation in transit depth due to the TLS effect, assuming a true transit depth of 3.4\% and a spot covering fraction of 20\%.}
    \label{fig:chromaticity}
\end{figure}

From the transit fitting results, we found a wavelength dependence in the transit depth $\delta \equiv k^2$ such that the transit depth is $\sim$7\% deeper in the $g$-band than in the $z$-band (Figure~\ref{fig:chromaticity}). A similar trend was reported in L23, and this study confirms their finding with the first high-precision, simultaneous multi-band transit observations.

This chromaticity could be caused either by the TLS effect, which is caused by unocculted spots, or by wavelength-dependent opacity in the planetary atmosphere. {The TLS effect refers to the alteration of the observed transit depth from the nominal value that would be measured in the absence of stellar surface inhomogeneity. Following the formulation of Equation 1 in \citet{Rackham2018}, and rewriting it for each photometric band $B$, the observed transit depth $D_{B,\,\rm obs}$ can be expressed as:
\begin{equation}
D_{B,\,\rm obs} = \frac{D_B}{1 - f_{\rm spot}(1 - c_B)},
\end{equation}
where $D_B$ is the nominal transit depth, $c_B$ is the spot contrast, and $f_{\rm spot}$ is the spot covering fraction, i.e., the fraction of the projected stellar disk covered by the spot.}
Assuming the best-fit values of photosphere and spot temperatures, we find that the observed chromaticity is consistent with a true transit depth of 3.4\% and a spot covering fraction of 20\%. However, the \texttt{fleck} model intrinsically accounts for the TLS effect by calculating the flux blocked by the planet relative to the flux from the modeled stellar surface. As the model assumes simple circular spots, it is plausible that more extended spot regions exist outside the transit chord. A more comprehensive treatment that simultaneously models spot-crossing and the TLS effect, using higher-precision, multi-wavelength light curves, would allow for a more rigorous evaluation of this chromaticity. We leave this for future work.

Alternatively, the observed chromaticity might be attributed to the planetary atmosphere. For instance, \citet{Thao2023} reported a transit depth in the optical that was twice as deep as in the infrared for the young planet K2-33b, interpreted as evidence for an extended, hazy atmosphere. TOI-3884b also has a relatively large radius for its level of stellar irradiation (A22, L23), placing it in a sparsely populated region of the planet radius–insolation diagram, which could suggest the presence of an extended atmosphere. However, based on the available data, it remains uncertain whether the observed chromaticity is atmospheric in origin. Spectroscopic transit observations, {such as those ongoing with the James Webb Space Telescope \citep[JWST;][]{Gardner2006} Cycle 3}\footnote{ID. 5799 and 5863, \url{https://www.stsci.edu/jwst/science-execution/approved-programs/general-observers/cycle-3-go}} will be crucial for disentangling these scenarios and retrieving the atmospheric composition of the planet.

\subsection{Long-term stability of the spot}\label{sec:dis-longterm}
Our transit observations and monitoring observations were conducted with a gap of 254 days between the last observed transit and the start of the monitoring campaign. During this period, it is possible that the shape and position of the spot changed. Therefore, we did not perform a simultaneous fitting of the two datasets. 

\begin{figure*}
    \centering
\includegraphics[width=0.8\linewidth]{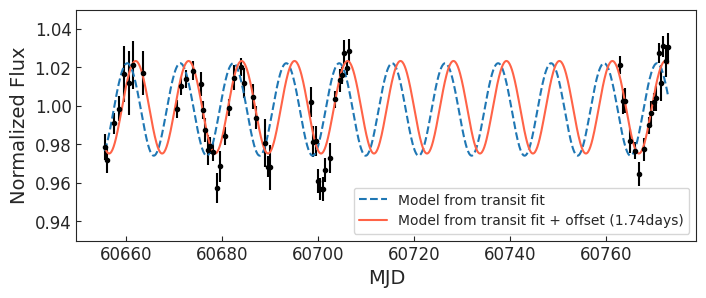}
    \caption{The rotational modulation light curve of TOI-3884 observed with Sinistro, overlaid with the model light curve generated using the best-fit parameters from spot-crossing analyses. The blue dashed line shows the forward model without any adjustment, while the red solid line represents the model after applying a time offset optimized to best match the observed modulation.}
    \label{fig:longterm}
\end{figure*}

However, we were interested in assessing how consistent the parameters obtained from the spot-crossing transit observations are with the modulation results. To examine this, Figure~\ref{fig:longterm} shows the modulation light curve observed during the monitoring campaign, overlaid with the forward-modeled light curve generated using the best-fit parameters from the transit analysis. The model was generated using \fleck to ensure consistency in the analysis. Note that, since \fleck assumes a disk-like spot shape without accounting for the projection onto a spherical stellar surface, it may overestimate the modulation amplitude. We compared the modulation model generated by \fleck with those obtained using a model that more accurately maps spots onto a spherical surface of the star, such as \texttt{macula} \citep{Kipping2012} as in \citet{Ikuta2023}, and it showed that \texttt{fleck} overestimates the modulation amplitudes by less than $1\%$, while the shapes of the modulation features remain similar.

{As shown in Figure~\ref{fig:longterm}, the model qualitatively reproduces the observed light curve. The amplitude of the modulation is $\sim$4.5\%, consistent with the observations. Additionally, we found that shifting the model by 1.74 days improves the phase alignment with the observed light curve. This offset could be interpreted as a change in the spot’s longitude and/or latitude. However, given the uncertainty in the stellar rotation period ($P_{\rm rot} \sim 0.05$ days), this level of offset can also be explained by error propagation.

Focusing on the shape of the modulation, the model generally agrees with the observed data but fails to reproduce the lowest flux levels. This is because the observed light curve exhibits a more sharply peaked shape than the smooth sinusoidal variation in the model. Such a feature can be better reproduced by introducing an additional spot at lower latitudes or by allowing for more complex spot geometries. Nonetheless, retrieving a unique spot configuration from single-band photometric monitoring is highly degenerate, and given the $>250$-day gap between the monitoring and transit observations, we refrain from further speculation on the spot distribution during the monitoring epoch.

Taken together, although minor differences in the modulation shape are present, we conclude that the overall spot properties on TOI-3884 remained relatively stable over a timescale of more than 200 days.}

On longer timescales (i.e., $>$1 year), variations in the spot’s configuration are suggested by the difference in modulation amplitude between the TESS light curves and our observations, as well as variability in the ZTF light curve modulation (Appendix~\ref{ap:ZTF}). Further long-term monitoring observations would be valuable to constrain the timescales over which spot properties evolve.

\subsection{Geometry of the TOI-3884 System}

Our analysis confirms the previously suggested {nearly polar-orbit} configuration of TOI-3884\,b and updates the relevant system parameters. The derived stellar inclination is {$i_\star = 139.9_{-2.0}^{+1.2}$ deg and the projected obliquity is $\lambda = 41.0_{-9.0}^{+3.7}$ deg. This result, when ignoring the direction of stellar rotation, corresponds to a solution with an absolute stellar inclination of $\sim$40\,deg, and} is therefore closer to the results of A22, than those of L23 (Table~\ref{tab:comparison}). If a stellar rotation period of 11.5 days is adopted instead, the solution becomes more consistent with that of L23. Distinguishing between these degenerate solutions would have been difficult without an accurate measurement of the stellar rotation period.

Using the derived values of $i_\star$, $i_{\rm orb}$, and $\lambda$, we calculate the true obliquity {$\Psi=61.9_{-5.6}^{+2.3}$} deg following the geometric relation {\citep[see][]{Fabrycky2009}}:

\begin{equation}
    \cos\Psi = \cos i_\star\cos i_{\rm orb}+\sin i_\star\sin i_{\rm orb}\cos\lambda \,\,.
\end{equation}

\noindent This places TOI-3884\,b among the small population of Neptune-sized planets known to have {close to polar} orbits \citep[e.g.,][]{Louden2024}.

\begin{figure}[ht]
    \centering
\includegraphics[width=1\linewidth]{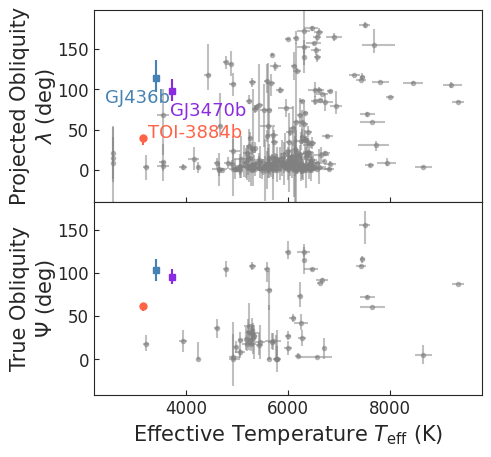}
    \caption{Distribution of measured stellar obliquities as a function of stellar effective temperature $T_{\rm eff}$. Data for systems other than TOI-3884b are taken from TEPCat \citep{Southworth2011}. TOI-3884b is highlighted in red. Systems with similar $T_{\rm eff}$ and obliquity to TOI-3884b, namely GJ~436b and GJ~3470b, are also highlighted.}
    \label{fig:obliquity}
\end{figure}

Figure~\ref{fig:obliquity} presents the distribution of obliquities for systems where $\lambda$ or $\Psi$ has been measured. Planets orbiting M dwarfs with similar obliquities include GJ~436b \citep{Bourrier2022} and GJ~3470b \citep{Stefansson2022}, both of which were confirmed to have polar orbits through Rossiter–McLaughlin (RM) effect measurements. However, their host stars are earlier-type and more slowly rotating than TOI-3884, and it remains unclear whether the same mechanisms are responsible for their {orbital misalignment.}

Because RM measurement for planets around late- to mid-type M dwarfs is particularly challenging due to their small $v \sin i_\star$ values and faintness, the current obliquity sample is biased toward young and rapidly rotating stars. In all of the few such young systems with RM measurements, aligned orbits have been observed, {including K2-25\,b \citep{Stefansson2020, Gaidos2020} and K2-33\,b \citep{Hirano2024}}, in contrast to TOI-3884\,b.

Recent studies have suggested that polar-orbit planets may exist at statistically significant rates, prompting discussion about their formation and stability \citep[e.g.,][]{Albrecht2021, Louden2024}. Many of the proposed mechanisms--such as disk-driven secular resonance \citep{Petrovich2020}--require the presence of an outer planet or stellar companion to excite the inner planet into a polar orbit. However, like GJ~436 and GJ~3470, TOI-3884 shows no signs of additional planets or stellar companions (A22, L23). Further radial velocity and astrometric follow-up will be important for determining the planet's mass and eccentricity, as well as for further searching for outer companions, all of which are crucial for understanding the system's dynamical evolution.

In addition to obliquity, we also find differences from previous studies in other transit parameters: semi-major axis $a/R_\star$ and impact parameter $b$. In particular, the impact parameter is found to be $0.40\pm0.02$, significantly higher than the near-zero value previously measured. This result persisted even when Gaussian priors based on previous studies were directly applied to $a/R_\star$ and $b$ in our transit fitting. While these two parameters are typically degenerate in transit fitting, the geometric constraints provided by spot-crossing events likely helped to break this degeneracy. However, it is also possible that the simplicity of the spot-crossing model introduced systematic biases, and this should be kept in mind when interpreting the results.

\section{Conclusion} \label{sec:conclusion}

Spot-crossing transits offer a unique opportunity to directly investigate spot temperature, size, and distribution. We conducted multi-epoch, multi-band photometry of TOI-3884\,b's transits using the MuSCAT instruments and optical photometric monitoring of the star TOI-3884 using LCO 1m telescopes/Sinistro.

Our analysis of these datasets yielded the following results:
\begin{itemize}
    \item We detected the stellar rotation signal of TOI-3884 for the first time from ground-based monitoring. The star has a rotation period of $11.043_{-0.053}^{+0.054}$ days, and the modulation amplitude was $\sim$5\% in $r$-band during the 2024–2025 winter season.
    
    \item The high-precision light curves obtained with the MuSCAT series allowed us to detect temporal variations in the shapes of the spot-crossing signals. {This provides the first evidence that the spot is offset from the stellar pole, such that its projected position along the transit chord varies with stellar rotation.}
    
    \item From our transit fitting, we derive a stellar inclination of {$i_\star\sim140$~deg, a projected obliquity of $\lambda\sim41$~deg, and a true obliquity of $\Psi\sim 62$~deg,} confirming the planet’s {highly misaligned} configuration. 

    \item Simultaneous multi-band photometry robustly confirmed that the spot temperature is $\sim$200K lower than the photosphere, which is consistent with previously measured spot temperatures on M dwarfs.

    \item The transit depth shows chromaticity, which can be explained by the effect of the unocculted portion of the spots or a possible planetary atmosphere. Spectroscopic transit observations are needed to distinguish these scenarios. 
    
    \item The absence of spot-crossing variability in the TESS light curves suggests that the spot may have been located closer to the stellar pole during the TESS observation epoch. ZTF data also indicate that spot properties have likely evolved over a timescale of several years.
\end{itemize}

Besides these, one of the important implications of this study is that TOI-3884 exhibits a large polar spot despite not being an ultra-fast rotator. This raises the possibility that polar spots may be common among field M dwarfs. In the case of TOI-3884, the transiting planet's polar orbit allowed the spot properties to be measured via spot-crossing events. In systems with aligned orbits, such constraints on polar spots would be impossible to obtain. Especially in the case of transmission spectroscopy for such systems, careful consideration of spots outside the transit chord is required, as these spots would cause the TLS effect. 

Nevertheless, several aspects remain to be explored in future work. In our analysis, we adopted a simple spot model, assuming a single circular spot, in order to avoid overfitting and to characterize the general features of the spot and system geometry. Future work should include more flexible spot models and account for faculae. Such analyses will require even higher-precision data. In particular, multi-wavelength transit observations combined with simultaneous stellar rotation monitoring would improve robustness. As discussed in Section~\ref{sec:dis-longterm}, long-term photometric monitoring would also help reveal temporal variations in spot properties. Due to its unusual geometric configuration, TOI-3884 may be uniquely suited to provide crucial insights into stellar surface inhomogeneities and magnetic activity in M dwarfs at large.

Finally, TOI-3884 is interesting not only for its spot properties and system geometry, but also as a target for atmospheric characterization. As re-confirmed by this work, TOI-3884\,b is a $\sim$6.4\,$R_\oplus$ planet transiting a mid-M dwarf with $T_{\rm eff} \sim 3150$\,K, yielding a relatively large planet-to-star radius ratio. The detected transit depth chromaticity may potentially be attributed to an extended planetary atmosphere. Ongoing and future spectroscopic observations are expected to reveal the detailed atmospheric composition of TOI-3884\,b. However, in order to accurately recover atmospheric signals from transmission spectra, it is essential to precisely estimate the influence of spots. Our results provide critical context for the interpretation of these spectroscopic data.

\begin{acknowledgments}
{After posting this manuscript on arXiv, we realized that the direction of the stellar inclination vector had been incorrectly defined. This affected several inferred parameters and descriptions, which have since been updated accordingly. We are deeply grateful to Dr. Patrick Tamburo for pointing out this important issue.

We also thank the anonymous referee for their insightful suggestions, and the editor for efficiently handling the manuscript.}

MM is supported by JSPS Grant-in-Aid for JSPS Fellows Grant Number JP24KJ0241, and 
JSPS KAKENHI Grant Number JP24K17083. 
This work is also supported by JSPS KAKENHI Grant Number JP24H00017, JP24K00689, JP25K17450, JP24K17082, JP24H00248, and JSPS Bilateral Program Number JPJSBP120249910.
This work is partly supported by JST SPRING, Grant Number JPMJSP2108.

This work makes use of observations made with the MuSCAT3/4 instruments, developed by the Astrobiology Center (ABC) in Japan, the University of Tokyo, and Las Cumbres Observatory (LCOGT). MuSCAT3 was developed with financial support by JSPS KAKENHI (JP18H05439) and JST PRESTO (JPMJPR1775), and is located at the Faulkes Telescope North on Maui, HI (USA), operated by LCOGT. MuSCAT4 was developed with financial support provided by the Heising-Simons Foundation (grant 2022-3611), JST grant number JPMJCR1761, and the ABC in Japan, and is located at the Faulkes Telescope South at Siding Spring Observatory (Australia), operated by LCOGT.

We acknowledge the use of public TESS data from pipelines at the TESS Science Office and at the TESS Science Processing Operations Center. Funding for the TESS mission is provided by NASA's Science Mission Directorate.  
All the TESS data used in this paper can be found in MAST: \dataset[10.17909/t9-nmc8-f686]{http://dx.doi.org/10.17909/t9-nmc8-f686}.

This work makes use of observations obtained with the Samuel Oschin Telescope 48-inch and the 60-inch Telescope at the Palomar Observatory as part of the Zwicky Transient Facility project. ZTF is supported by the National Science Foundation under Grant No. AST-2034437 and a collaboration including Caltech, IPAC, the Weizmann Institute for Science, the Oskar Klein Center at Stockholm University, the University of Maryland, Deutsches Elektronen-Synchrotron and Humboldt University, the TANGO Consortium of Taiwan, the University of Wisconsin at Milwaukee, Trinity College Dublin, Lawrence Livermore National Laboratories, and IN2P3, France. Operations are conducted by COO, IPAC, and UW.

This work has made use of data from the European Space Agency (ESA) mission {\it Gaia} (\url{https://www.cosmos.esa.int/gaia}), processed by the {\it Gaia} Data Processing and Analysis Consortium (DPAC, \url{https://www.cosmos.esa.int/web/gaia/dpac/consortium}). Funding for the DPAC has been provided by national institutions, in particular the institutions participating in the {\it Gaia} Multilateral Agreement.

The authors used ChatGPT (version GPT-4), a language model developed by OpenAI, to help improve the grammar and clarity of the manuscript.

\end{acknowledgments}

\vspace{5mm}
\facilities{LCOGT (Sinistro), FTN (MuSCAT3), FTS (MuSCAT4), TESS, PO:1.2m (ZTF), Exoplanet Archive, Gaia.} 

\software{
\AstroImageJ \citep{Collins2017}, 
\texttt{Astropy} \citep{astropy2013,astropy2018},
\Photutils \citep{photutils}, 
\texttt{Astroquery} \citep{Astroquery2019},
\BANZAI \citep{McCully2018}, 
\batman \citep{Kreidberg2015}, 
\celerite \citep{Foreman-Mackey2017F}, 
\texttt{corner} \citep{Foreman-Mackey2016},
\emcee \citep{Foreman-Mackey2013}, 
\fleck \citep{Morris2020}, 
\texttt{Jupyter} \citep{jupyter2016},
\lightkurve \citep{lk2018},
\texttt{Matplotlib} \citep{matplotlib2007},
\texttt{Numpy} \citep{numpy2020},
\texttt{pandas} \citep{pandas2022},
\PyPeriod \citep{pya}, 
\ldtk \citep{Parviainen2015,Husser2013}, 
\texttt{scipy} \citep{scipy2020}.
}

\appendix

\section{Comparison of posterior distributions from transit fitting}\label{ap:comp}

As described in Section~\ref{sec:analyses}, the posterior distributions obtained from the transit fitting show a bimodal structure. This bimodality is particularly evident in parameters such as the stellar inclination $i_\star$, the projected obliquity $\lambda$, and the spot latitude and longitude, while the other parameters exhibit approximately Gaussian distributions.

\begin{figure}[p]
    \centering
\includegraphics[width=1\linewidth]{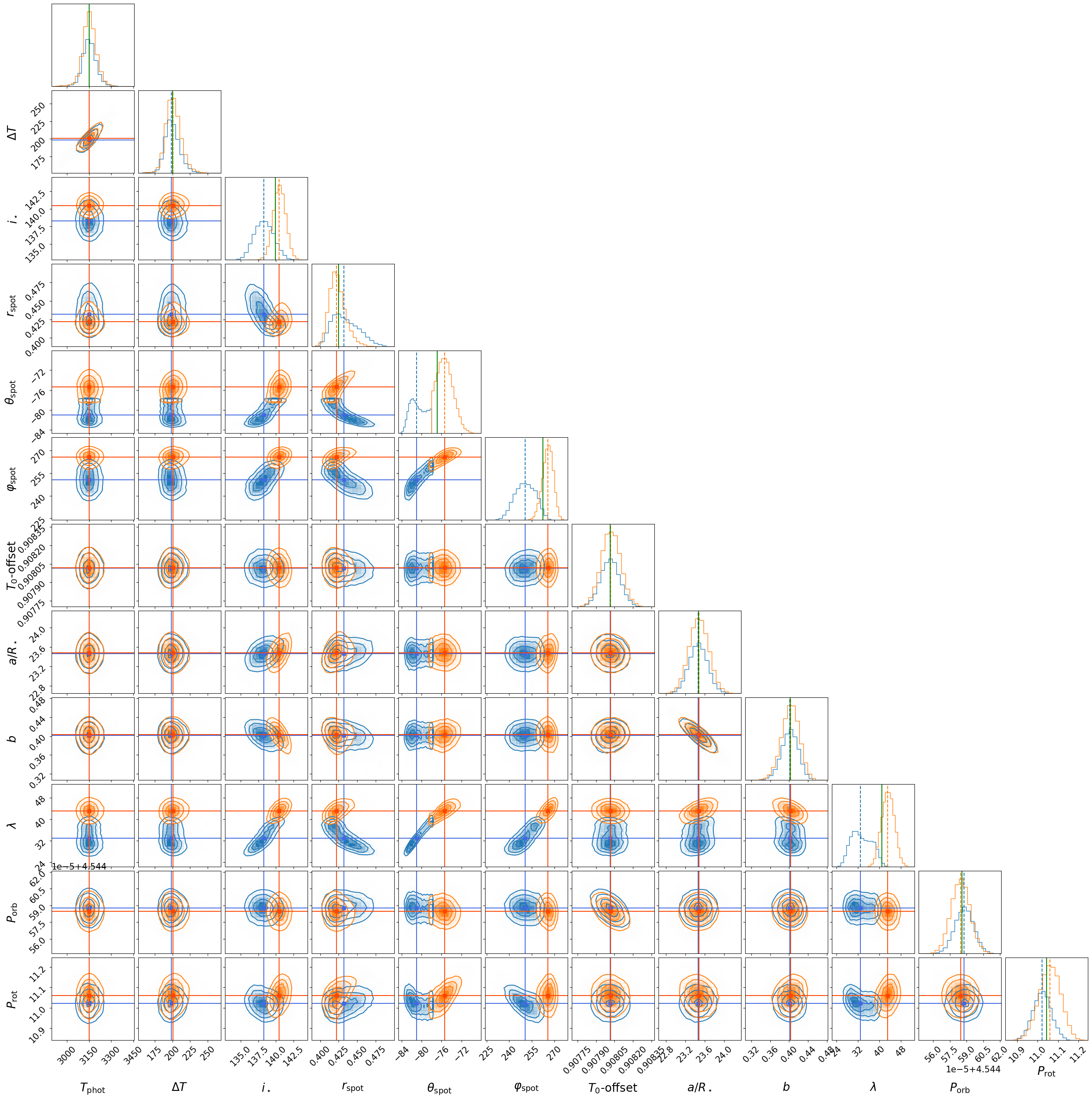}
    \caption{Corner plot showing the posterior distributions of the fitted parameters from the transit analysis. The plot includes a subset of the posteriors, omitting the 12 additional parameters ($k$, $q_1$, $q_2$ for each band) that were also sampled. {Blue and orange contours represent posterior samples from group 1 (spot latitude $\theta_{\rm spot}\leq-78$\,deg) and group 2 ($>-78$\,deg), respectively. The apparent overlap in the $\theta_{\rm spot}$ contours is due to smoothing in the visualization and does not indicate actual mixing.} The blue and orange lines indicate the best-fit parameter values for each group, respectively. The {green solid line in each histogram marks the best-fit value from the full posterior.}}
    \label{fig:corner}
\end{figure}

To investigate these two peaks, we divided the converged MCMC samples into two groups based on the spot latitude: group 1 includes samples with a spot latitude {smaller than -78 deg, and group 2 includes those with higher values}. Figure~\ref{fig:corner} presents the corner plots of the posterior distributions for each group. From each group, we extracted the median value of each parameter as the best-fit solution.

\begin{figure}[ht]
    \centering
\includegraphics[width=0.9\linewidth]{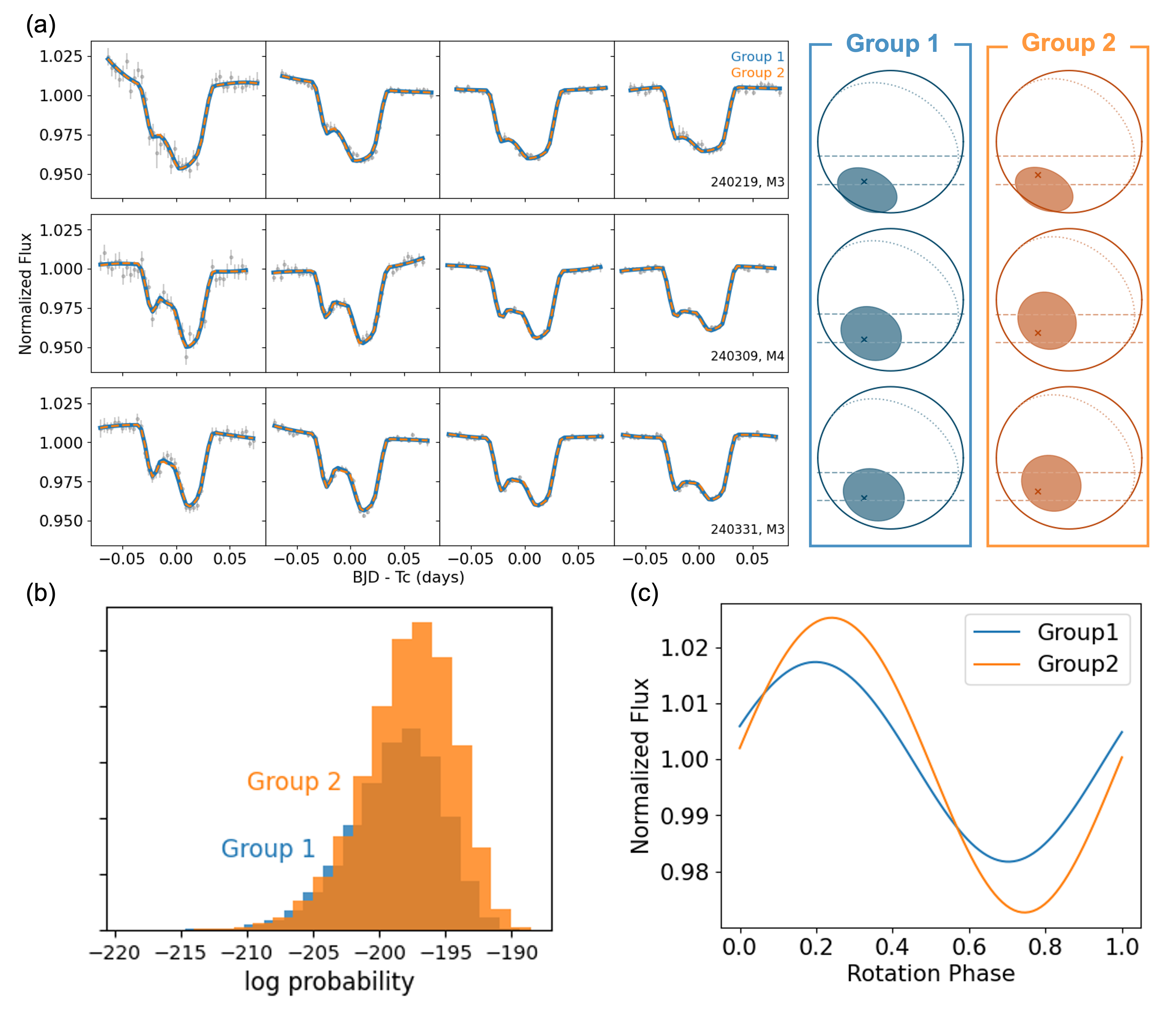}
    \caption{{Comparison of the two best-fit models from group 1 (blue) and group 2 (orange).  (a) Model light curves for each group, overlaid on the same observed data as shown in Figure~\ref{fig:transit_best}, plotted in gray here for clarity.
    The two models are nearly indistinguishable. The right panels show the corresponding spot configurations at the time of each transit. (b) Distribution of log-probability values from the transit fits.  (c) Predicted rotational modulation as a function of stellar rotation phase.}}
\label{fig:spot_hikaku}
\end{figure}

We find that the spot-crossing transit models generated from the best-fit parameters of the two groups are visually indistinguishable in the transit light curves and show similar log probability values (Figure~\ref{fig:spot_hikaku}). The difference between the two solutions arises primarily from whether the circular spot lies slightly above or below the transit chord during the second and third transits -- a distinction that cannot be resolved with the current photometric data alone.

These two solutions show slightly different amplitudes and phases in rotational modulation (Figure~\ref{fig:spot_hikaku}). Therefore, discriminating between these two configurations would require high-precision, multi-epoch transit observations conducted simultaneously with rotational monitoring. Nevertheless, the two distributions agree within approximately 2$\sigma$ for all parameters, and do not significantly affect the overall interpretation of the system. The final parameter values reported in Table~\ref{tab:retrieved_params} are based on the median and 1$\sigma$ credible intervals computed from the full set of converged posterior samples.

\newpage
\section{Period search in the ZTF light curves}\label{ap:ZTF}

The Zwicky Transient Facility (ZTF) is a photometric survey instrument on the Samuel Oschin 1.2m telescope at Palomar Observatory in California, which conducts a full sweep of the northern sky every two days \citep{ztf1,ztf2}. To investigate possible rotational signals, we analyzed archival photometry of TOI-3884 in the $g$- and $r$-bands from ZTF Data Release 22 \citep{https://doi.org/10.26131/irsa598}. We removed data points with quality warning flags (``catflags'' in the ZTF catalog) and those with zero-point magnitudes more than $3\sigma$ below the median, to filter out nights with poor observing conditions. The magnitude was corrected for the second-order extinction effect using a linear function of color coefficients (``clrcoeff'' in the ZTF catalog). Finally, the magnitude was converted to flux, from which we produced a normalized light curve.  

\begin{figure}[ht]
    \centering
\includegraphics[width=1\linewidth]{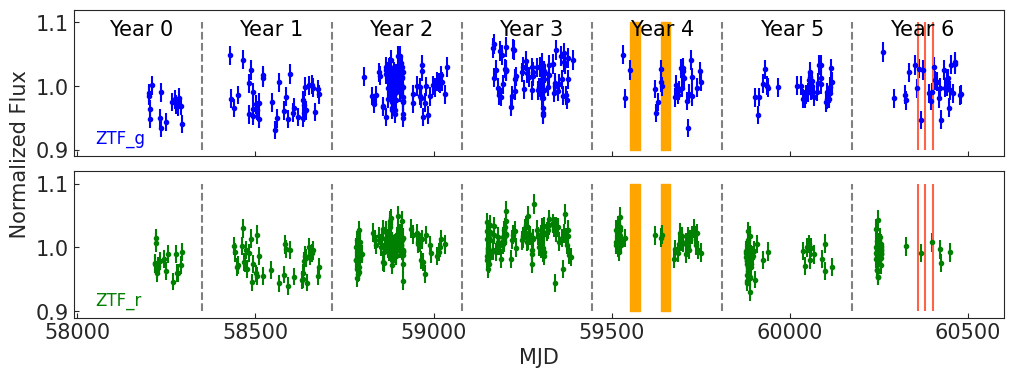}
    \caption{Light curves of TOI-3884 observed with ZTF in the $g$-band (blue, upper figure) and $r$-band (green, lower figure). The gray dashed lines indicate the boundaries between different observing seasons (Year 0 to Year 6). The orange-shaded regions mark the periods of TESS observations in Sectors 46 and 49, from left to right. The red lines indicate the timings of the three transit observations conducted with MuSCAT3 and 4.}
    \label{fig:ZTF_all}
\end{figure}

Figure~\ref{fig:ZTF_all} shows the light curves of TOI-3884 in the ZTF $g$- and $r$-bands, taken over a six-year period from March 2018 to July 2024. For analysis, we divided the data into seasons, labeling them as Years 0-6. The TESS observations in Sectors 46 and 49 occurred in Year 4, while the MuSCAT3/4 transit observations were conducted in Year 6. The Sinistro monitoring observations took place the following year, i.e., Year 7.  

\begin{figure}[ht]
    \centering
\includegraphics[width=1\linewidth]{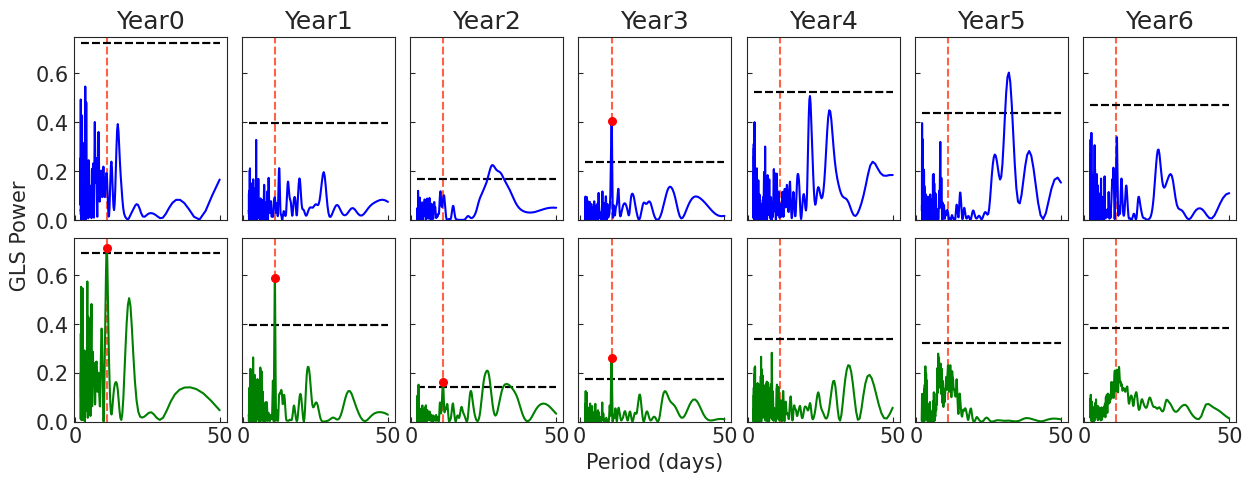}
    \caption{GLS periodograms of TOI-3884's ZTF light curves for each band and each year. The upper panels (blue) correspond to the $g$-band, while the lower panels (green) correspond to the $r$-band. The black horizontal dashed lines indicate the FAP = 1\% threshold. The red vertical dashed lines mark the period of 11.04 days, and red dots indicate cases where peaks are detected near this period with FAP $<1$\%.}
    \label{fig:ZTF_gls}
\end{figure}

For each year, we performed a period search on the light curves using the GLS periodogram in the range of 2 to 50 days (Figure~\ref{fig:ZTF_gls}). In particular, the $r$-band exhibit a periodic signal around $P \sim 11$ days in multiple years with a false alarm probability (FAP) below 1\%. This supports the stellar rotation period independently determined from the Sinistro observations. Other signals exceeding the detection threshold, such as 26 days, 28 days, and 32 days, were also identified, but they were not consistently detected in other years.  

\begin{figure}[ht]
    \centering
\includegraphics[width=1\linewidth]{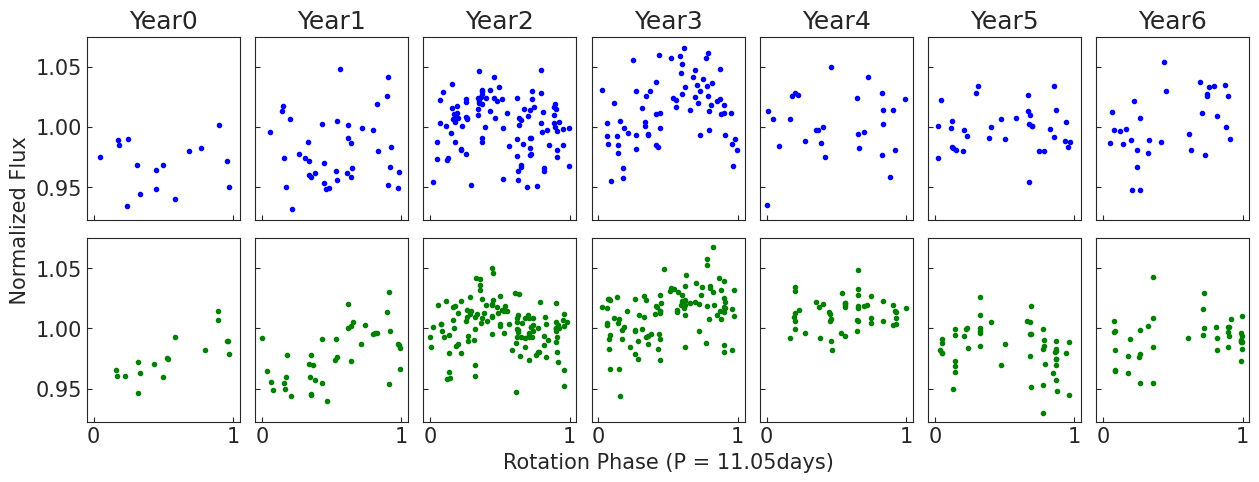}
    \caption{Phase-folded ZTF light curves of TOI-3884 with a period of 11.04 days, for each band and each year. The panel arrangement follows the same order as in Figure~\ref{fig:ZTF_gls}.}
    \label{fig:ZTF_folded}
\end{figure}

Furthermore, we phase-folded the light curves for each year using the 11.04-day period (Figure~\ref{fig:ZTF_folded}). The results indicate that the amplitude of variability changes over time, with some years showing stronger modulation than others. The apparent phase shifts could be due to uncertainties in the period determination or spot evolution. Although the number of data points is limited, the modulation amplitude during Year 4, when TESS observations were conducted, appears to be smaller than in other years. This suggests that the stellar activity level may have been relatively low during that period.

\newpage
\section{Reanalysis of TESS light curves}\label{ap:TESS}

TOI-3884 was observed by TESS in Sector 46 (2021 December 2 to December 30) and Sector 49 (2022 February 26 to March 25). Using the \texttt{lightkurve} package \citep{lk2018}, we downloaded both the SAP light curves and the Pre-search Data Conditioning SAP (PDCSAP) light curves, which have long-term trends removed. 

To investigate the spot configuration during the TESS observations, we carried out transit fitting using the PDCSAP light curve. We identified four transits in each sector (eight in total) and extracted light curve segments spanning $\pm$0.15 days before and after each predicted transit center. Outliers deviating more than $3\sigma$ from the median flux were masked, and the light curves were binned into 5-min intervals.

The transit fitting was conducted using the same approach described in Section~\ref{sec:analyses}, but with a total of 15 free parameters, due to the availability of only one photometric band (TESS). Orbital parameters that are not expected to vary with wavelength or epoch, including $T_{\rm phot}$, which is poorly constrained from a single band, were assigned Gaussian priors based on the MuSCAT fit results (Table~\ref{tab:retrieved_params}). For $\Delta T$, we adopted a Gaussian prior centered at 200\,K with a relatively broad standard deviation of 100\,K to account for the possible change in spot temperature. 

\begin{figure}
    \centering
    \includegraphics[width=1\linewidth]{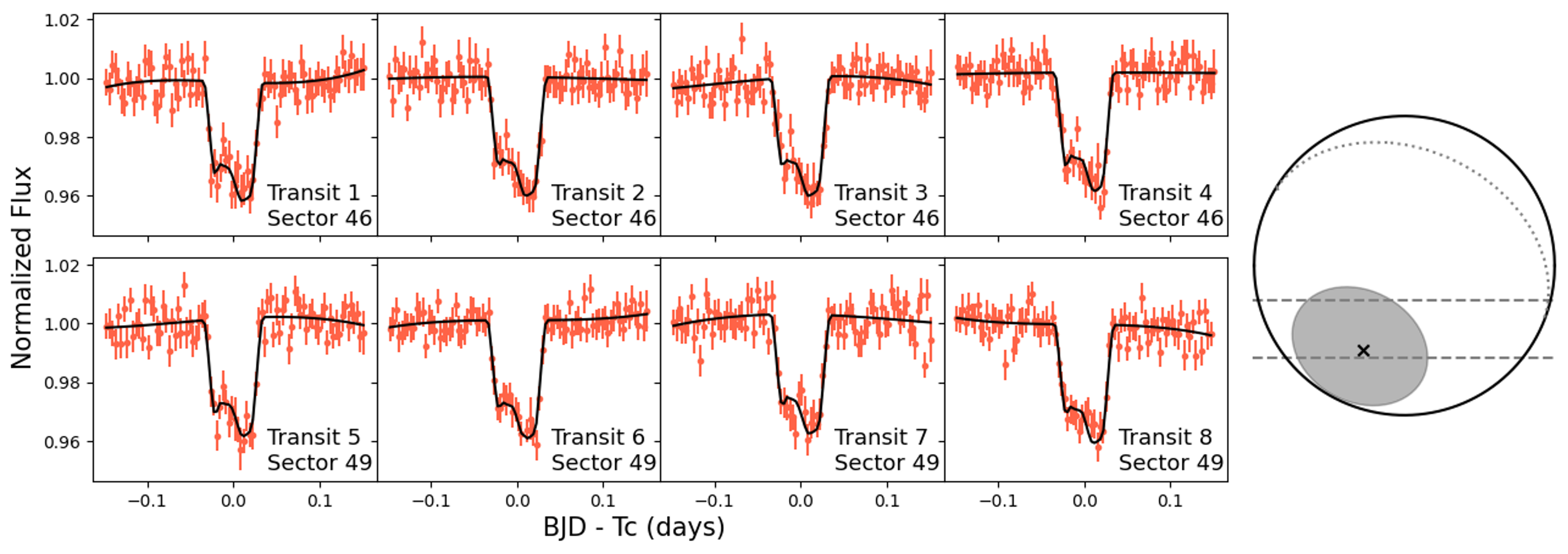}
    \caption{Light curves of the eight transits observed by TESS, overlaid with the best-fit transit models. The right panel shows the corresponding spot configuration on the stellar surface at the time of the TESS observations, as derived from the best-fit parameters. Since the spot latitude is close to the pole, the spot appears in almost the same position in all transits.}
    \label{fig:tess_transit}
\end{figure}

The priors and posteriors for the TESS transit fit are summarized in Table~\ref{tab:tess_params}, and the best-fit model light curves are shown in Figure~\ref{fig:tess_transit}. These results suggest a spot latitude of {$\sim$ -88 deg}, consistent with a nearly exact polar spot and the absence of rotational modulation in the TESS light curves.

\begin{table*}
\centering
\caption{Parameters of TOI-3884 system from TESS observations. \label{tab:tess_params}}
\begin{tabular}{llcc}\hline\hline
Name & Parameter [Units] & Prior & Derived Value \\
\hline
\multicolumn{4}{l}{\it $\diamond$ Sampled parameters} \\
Photosphere temperature & $T_{\rm phot}$ [K] & $\mathcal{U}$(2700, 5000) $\times$ $\mathcal{N}$(3151, 45) & $3153 \pm 44$\\
Spot temperature difference & $\Delta T$ [K] & $\mathcal{U}$(0, 1000)$\times$ $\mathcal{N}$(200, 100) & $161 \pm 19$\\
Spot radius & $r_{\rm spot} [R_\star]$ & $\mathcal{U}$(0, 0.5)& $0.472_{-0.037}^{+0.020}$\\
Spot latitude & $\theta_{\rm spot}$ [deg] & $\mathcal{U}$(-90, 90)& {$-88.1_{-1.3}^{+1.9}$}\\
Spot longitude & $\varphi_{\rm spot}$ [deg] & $\mathcal{U}$(0, 360)& {$223_{-91}^{+74}$}\\\\
Transit center & $T_0 - 2459556$ [BJD]& $\mathcal{N}$(0.516690, 0.000078) & $0.516705 \pm 0.000075$\\
Orbital period & $P_{\rm orb}$ [days] & $\mathcal{N}$(4.5445858, 0.0000086)& $4.5445867 \pm 0.0000074$ \\
Semimajor axis & $a/R_\star$ & $\mathcal{U}$(0, 100)  $\times$  $\mathcal{N}$(23.47, 0.20)& $23.60 \pm 0.17$\\
Impact parameter & $b$ & $\mathcal{U}$(0, 1)  $\times$  $\mathcal{N}$(0.402, 0.019)& $0.427 \pm 0.015$\\
Planetary radius & $R_p/R_\star$~(TESS)& $\mathcal{U}$(0, 1) $\times$ $\mathcal{N}$(0.197, 0.002) & $0.1921 \pm 0.0015$\\
Rotation period & $P_{\rm rot}$ [days] & $\mathcal{U}$(10.8, 11.3) $\times$ $\mathcal{N}$(11.043, 0.054)& $11.041 \pm 0.055$\\
Stellar inclination & $i_\star$ [deg] &  $\mathcal{U}$(0, 90)$\times$ $\mathcal{N}${(139.9, 2.0)}& {$141.1 \pm 1.9$}\\
Projected obliquity & $\lambda$ [deg] & $\mathcal{U}$(-180, 180)$\times$ $\mathcal{N}${(41.0, 9.0)}& {$26.2_{-4.2}^{+5.6}$}\\
limb-darkening & $q_1$~(TESS) & $\mathcal{U}$(0, 1) $\times$ $\mathcal{N}$(0.30, 0.10) & $0.061_{-0.049}^{+0.089}$\\
coefficients & $q_2$~(TESS) & $\mathcal{U}$(0, 1) $\times$  $\mathcal{N}$(0.29, 0.05) & $0.288_{-0.049}^{+0.050}$ \\
\hline\hline
\multicolumn{4}{l}{$^\ast$ $\mathcal{U}$(A, B) indicates the uniform prior from A to B, and $\mathcal{N}$(X, Y) indicates the Gaussian prior with the mean value X}\\ \multicolumn{4}{l}{~~and standard deviation Y.}\\
\end{tabular}
\end{table*}

\newpage
\bibliography{ref}{}

\begin{thebibliography}{}
\expandafter\ifx\csname natexlab\endcsname\relax\def\natexlab#1{#1}\fi
\providecommand{\url}[1]{\href{#1}{#1}}
\providecommand{\dodoi}[1]{doi:~\href{http://doi.org/#1}{\nolinkurl{#1}}}
\providecommand{\doeprint}[1]{\href{http://ascl.net/#1}{\nolinkurl{http://ascl.net/#1}}}
\providecommand{\doarXiv}[1]{\href{https://arxiv.org/abs/#1}{\nolinkurl{https://arxiv.org/abs/#1}}}

\bibitem[{S.~H. {Albrecht} {et~al.}(2021){Albrecht}, {Marcussen}, {Winn}, {Dawson}, \& {Knudstrup}}]{Albrecht2021}
{Albrecht}, S.~H., {Marcussen}, M.~L., {Winn}, J.~N., {Dawson}, R.~I., \& {Knudstrup}, E. 2021, \bibinfo{title}{{A Preponderance of Perpendicular Planets},} \apjl, 916, L1, \dodoi{10.3847/2041-8213/ac0f03}

\bibitem[{F. {Allard}(2014){Allard}}]{Allard2014}
{Allard}, F. 2014, in IAU Symposium, Vol. 299, Exploring the Formation and Evolution of Planetary Systems, ed. M.~{Booth}, B.~C. {Matthews}, \& J.~R. {Graham}, 271--272, \dodoi{10.1017/S1743921313008545}

\bibitem[{J.~M. {Almenara} {et~al.}(2022){Almenara}, {Bonfils}, {Forveille}, {Astudillo-Defru}, {Ciardi}, {Schwarz}, {Collins}, {Cointepas}, {Lund}, {Bouchy}, {Charbonneau}, {D{\'\i}az}, {Delfosse}, {Kidwell}, {Kunimoto}, {Latham}, {Lissauer}, {Murgas}, {Ricker}, {Seager}, {Vezie}, \& {Watanabe}}]{Almenara2022}
{Almenara}, J.~M., {Bonfils}, X., {Forveille}, T., {et~al.} 2022, \bibinfo{title}{{TOI-3884 b: A rare 6-R$_{E}$ planet that transits a low-mass star with a giant and likely polar spot},} \aap, 667, L11, \dodoi{10.1051/0004-6361/202244791}

\bibitem[{R. {Angus} {et~al.}(2018){Angus}, {Morton}, {Aigrain}, {Foreman-Mackey}, \& {Rajpaul}}]{Angus2018}
{Angus}, R., {Morton}, T., {Aigrain}, S., {Foreman-Mackey}, D., \& {Rajpaul}, V. 2018, \bibinfo{title}{{Inferring probabilistic stellar rotation periods using Gaussian processes},} \mnras, 474, 2094, \dodoi{10.1093/mnras/stx2109}

\bibitem[{ {Astropy Collaboration} {et~al.}(2013){Astropy Collaboration}, {Robitaille}, {Tollerud}, {Greenfield}, {Droettboom}, {Bray}, {Aldcroft}, {Davis}, {Ginsburg}, {Price-Whelan}, {Kerzendorf}, {Conley}, {Crighton}, {Barbary}, {Muna}, {Ferguson}, {Grollier}, {Parikh}, {Nair}, {Unther}, {Deil}, {Woillez}, {Conseil}, {Kramer}, {Turner}, {Singer}, {Fox}, {Weaver}, {Zabalza}, {Edwards}, {Azalee Bostroem}, {Burke}, {Casey}, {Crawford}, {Dencheva}, {Ely}, {Jenness}, {Labrie}, {Lim}, {Pierfederici}, {Pontzen}, {Ptak}, {Refsdal}, {Servillat}, \& {Streicher}}]{astropy2013}
{Astropy Collaboration}, {Robitaille}, T.~P., {Tollerud}, E.~J., {et~al.} 2013, \bibinfo{title}{{Astropy: A community Python package for astronomy},} \aap, 558, A33, \dodoi{10.1051/0004-6361/201322068}

\bibitem[{ {Astropy Collaboration} {et~al.}(2018){Astropy Collaboration}, {Price-Whelan}, {Sip{\H{o}}cz}, {G{\"u}nther}, {Lim}, {Crawford}, {Conseil}, {Shupe}, {Craig}, {Dencheva}, {Ginsburg}, {VanderPlas}, {Bradley}, {P{\'e}rez-Su{\'a}rez}, {de Val-Borro}, {Aldcroft}, {Cruz}, {Robitaille}, {Tollerud}, {Ardelean}, {Babej}, {Bach}, {Bachetti}, {Bakanov}, {Bamford}, {Barentsen}, {Barmby}, {Baumbach}, {Berry}, {Biscani}, {Boquien}, {Bostroem}, {Bouma}, {Brammer}, {Bray}, {Breytenbach}, {Buddelmeijer}, {Burke}, {Calderone}, {Cano Rodr{\'\i}guez}, {Cara}, {Cardoso}, {Cheedella}, {Copin}, {Corrales}, {Crichton}, {D'Avella}, {Deil}, {Depagne}, {Dietrich}, {Donath}, {Droettboom}, {Earl}, {Erben}, {Fabbro}, {Ferreira}, {Finethy}, {Fox}, {Garrison}, {Gibbons}, {Goldstein}, {Gommers}, {Greco}, {Greenfield}, {Groener}, {Grollier}, {Hagen}, {Hirst}, {Homeier}, {Horton}, {Hosseinzadeh}, {Hu}, {Hunkeler}, {Ivezi{\'c}}, {Jain}, {Jenness}, {Kanarek}, {Kendrew}, {Kern}, {Kerzendorf}, {Khvalko}, {King}, {Kirkby}, {Kulkarni},
  {Kumar}, {Lee}, {Lenz}, {Littlefair}, {Ma}, {Macleod}, {Mastropietro}, {McCully}, {Montagnac}, {Morris}, {Mueller}, {Mumford}, {Muna}, {Murphy}, {Nelson}, {Nguyen}, {Ninan}, {N{\"o}the}, {Ogaz}, {Oh}, {Parejko}, {Parley}, {Pascual}, {Patil}, {Patil}, {Plunkett}, {Prochaska}, {Rastogi}, {Reddy Janga}, {Sabater}, {Sakurikar}, {Seifert}, {Sherbert}, {Sherwood-Taylor}, {Shih}, {Sick}, {Silbiger}, {Singanamalla}, {Singer}, {Sladen}, {Sooley}, {Sornarajah}, {Streicher}, {Teuben}, {Thomas}, {Tremblay}, {Turner}, {Terr{\'o}n}, {van Kerkwijk}, {de la Vega}, {Watkins}, {Weaver}, {Whitmore}, {Woillez}, {Zabalza}, \& {Astropy Contributors}}]{astropy2018}
{Astropy Collaboration}, {Price-Whelan}, A.~M., {Sip{\H{o}}cz}, B.~M., {et~al.} 2018, \bibinfo{title}{{The Astropy Project: Building an Open-science Project and Status of the v2.0 Core Package},} \aj, 156, 123, \dodoi{10.3847/1538-3881/aabc4f}

\bibitem[{E.~C. {Bellm} {et~al.}(2019){Bellm}, {Kulkarni}, {Graham}, {Dekany}, {Smith}, {Riddle}, {Masci}, {Helou}, {Prince}, {Adams}, {Barbarino}, {Barlow}, {Bauer}, {Beck}, {Belicki}, {Biswas}, {Blagorodnova}, {Bodewits}, {Bolin}, {Brinnel}, {Brooke}, {Bue}, {Bulla}, {Burruss}, {Cenko}, {Chang}, {Connolly}, {Coughlin}, {Cromer}, {Cunningham}, {De}, {Delacroix}, {Desai}, {Duev}, {Eadie}, {Farnham}, {Feeney}, {Feindt}, {Flynn}, {Franckowiak}, {Frederick}, {Fremling}, {Gal-Yam}, {Gezari}, {Giomi}, {Goldstein}, {Golkhou}, {Goobar}, {Groom}, {Hacopians}, {Hale}, {Henning}, {Ho}, {Hover}, {Howell}, {Hung}, {Huppenkothen}, {Imel}, {Ip}, {Ivezi{\'c}}, {Jackson}, {Jones}, {Juric}, {Kasliwal}, {Kaspi}, {Kaye}, {Kelley}, {Kowalski}, {Kramer}, {Kupfer}, {Landry}, {Laher}, {Lee}, {Lin}, {Lin}, {Lunnan}, {Giomi}, {Mahabal}, {Mao}, {Miller}, {Monkewitz}, {Murphy}, {Ngeow}, {Nordin}, {Nugent}, {Ofek}, {Patterson}, {Penprase}, {Porter}, {Rauch}, {Rebbapragada}, {Reiley}, {Rigault}, {Rodriguez}, {van Roestel}, {Rusholme},
  {van Santen}, {Schulze}, {Shupe}, {Singer}, {Soumagnac}, {Stein}, {Surace}, {Sollerman}, {Szkody}, {Taddia}, {Terek}, {Van Sistine}, {van Velzen}, {Vestrand}, {Walters}, {Ward}, {Ye}, {Yu}, {Yan}, \& {Zolkower}}]{ztf1}
{Bellm}, E.~C., {Kulkarni}, S.~R., {Graham}, M.~J., {et~al.} 2019, \bibinfo{title}{{The Zwicky Transient Facility: System Overview, Performance, and First Results},} \pasp, 131, 018002, \dodoi{10.1088/1538-3873/aaecbe}

\bibitem[{S.~V. {Berdyugina}(2005){Berdyugina}}]{Berdyugina2005}
{Berdyugina}, S.~V. 2005, \bibinfo{title}{{Starspots: A Key to the Stellar Dynamo},} Living Reviews in Solar Physics, 2, 8, \dodoi{10.12942/lrsp-2005-8}

\bibitem[{V. {Bourrier} {et~al.}(2022){Bourrier}, {Zapatero Osorio}, {Allart}, {Attia}, {Cretignier}, {Dumusque}, {Lovis}, {Adibekyan}, {Borsa}, {Figueira}, {Gonz{\'a}lez Hern{\'a}ndez}, {Mehner}, {Santos}, {Schmidt}, {Seidel}, {Sozzetti}, {Alibert}, {Casasayas-Barris}, {Ehrenreich}, {Lo Curto}, {Martins}, {Di Marcantonio}, {M{\'e}gevand}, {Nunes}, {Palle}, {Poretti}, \& {Sousa}}]{Bourrier2022}
{Bourrier}, V., {Zapatero Osorio}, M.~R., {Allart}, R., {et~al.} 2022, \bibinfo{title}{{The polar orbit of the warm Neptune GJ 436b seen with VLT/ESPRESSO},} \aap, 663, A160, \dodoi{10.1051/0004-6361/202142559}

\bibitem[{L. {Bradley} {et~al.}(2021){Bradley}, {Sip{\H{o}}cz}, {Robitaille}, {Tollerud}, {Vin{\'\i}cius}, {Deil}, {Barbary}, {Wilson}, {Busko}, {Donath}, {G{\"u}nther}, {Cara}, {Krachyon}, {Conseil}, {Bostroem}, {Droettboom}, {Bray}, {Lim}, {Andersen Bratholm}, {Barentsen}, {Craig}, {Rathi}, {Pascual}, {Perren}, {Georgiev}, {De Val-Borro}, {Kerzendorf}, {Bach}, {Quint}, \& {Souchereau}}]{photutils}
{Bradley}, L., {Sip{\H{o}}cz}, B., {Robitaille}, T., {et~al.} 2021, \bibinfo{title}{{astropy/photutils: 1.3.0},}, 1.3.0, Zenodo Zenodo, \dodoi{10.5281/zenodo.5796924}

\bibitem[{T.~M. {Brown} {et~al.}(2013){Brown}, {Baliber}, {Bianco}, {Bowman}, {Burleson}, {Conway}, {Crellin}, {Depagne}, {De Vera}, {Dilday}, {Dragomir}, {Dubberley}, {Eastman}, {Elphick}, {Falarski}, {Foale}, {Ford}, {Fulton}, {Garza}, {Gomez}, {Graham}, {Greene}, {Haldeman}, {Hawkins}, {Haworth}, {Haynes}, {Hidas}, {Hjelstrom}, {Howell}, {Hygelund}, {Lister}, {Lobdill}, {Martinez}, {Mullins}, {Norbury}, {Parrent}, {Paulson}, {Petry}, {Pickles}, {Posner}, {Rosing}, {Ross}, {Sand}, {Saunders}, {Shobbrook}, {Shporer}, {Street}, {Thomas}, {Tsapras}, {Tufts}, {Valenti}, {Vander Horst}, {Walker}, {White}, \& {Willis}}]{Brown2013}
{Brown}, T.~M., {Baliber}, N., {Bianco}, F.~B., {et~al.} 2013, \bibinfo{title}{{Las Cumbres Observatory Global Telescope Network},} \pasp, 125, 1031, \dodoi{10.1086/673168}

\bibitem[{K.~A. {Collins} {et~al.}(2017){Collins}, {Kielkopf}, {Stassun}, \& {Hessman}}]{Collins2017}
{Collins}, K.~A., {Kielkopf}, J.~F., {Stassun}, K.~G., \& {Hessman}, F.~V. 2017, \bibinfo{title}{{AstroImageJ: Image Processing and Photometric Extraction for Ultra-precise Astronomical Light Curves},} \aj, 153, 77, \dodoi{10.3847/1538-3881/153/2/77}

\bibitem[{S. {Czesla} {et~al.}(2019){Czesla}, {Schr{\"o}ter}, {Schneider}, {Huber}, {Pfeifer}, {Andreasen}, \& {Zechmeister}}]{pya}
{Czesla}, S., {Schr{\"o}ter}, S., {Schneider}, C.~P., {et~al.} 2019, \bibinfo{title}{{PyA: Python astronomy-related packages},} \doeprint{1906.010}

\bibitem[{J.~R.~A. {Davenport} {et~al.}(2015){Davenport}, {Hebb}, \& {Hawley}}]{Davenport2015}
{Davenport}, J. R.~A., {Hebb}, L., \& {Hawley}, S.~L. 2015, \bibinfo{title}{{Detecting Differential Rotation and Starspot Evolution on the M Dwarf GJ 1243 with Kepler},} \apj, 806, 212, \dodoi{10.1088/0004-637X/806/2/212}

\bibitem[{D.~C. {Fabrycky} \& J.~N. {Winn}(2009){Fabrycky} \& {Winn}}]{Fabrycky2009}
{Fabrycky}, D.~C., \& {Winn}, J.~N. 2009, \bibinfo{title}{{Exoplanetary Spin-Orbit Alignment: Results from the Ensemble of Rossiter-McLaughlin Observations},} \apj, 696, 1230, \dodoi{10.1088/0004-637X/696/2/1230}

\bibitem[{D. {Foreman-Mackey}(2016){Foreman-Mackey}}]{Foreman-Mackey2016}
{Foreman-Mackey}, D. 2016, \bibinfo{title}{{corner.py: Scatterplot matrices in Python},} The Journal of Open Source Software, 1, 24, \dodoi{10.21105/joss.00024}

\bibitem[{D. {Foreman-Mackey} {et~al.}(2017){Foreman-Mackey}, {Agol}, {Ambikasaran}, \& {Angus}}]{Foreman-Mackey2017F}
{Foreman-Mackey}, D., {Agol}, E., {Ambikasaran}, S., \& {Angus}, R. 2017, \bibinfo{title}{{Fast and Scalable Gaussian Process Modeling with Applications to Astronomical Time Series},} \aj, 154, 220, \dodoi{10.3847/1538-3881/aa9332}

\bibitem[{D. {Foreman-Mackey} {et~al.}(2013){Foreman-Mackey}, {Hogg}, {Lang}, \& {Goodman}}]{Foreman-Mackey2013}
{Foreman-Mackey}, D., {Hogg}, D.~W., {Lang}, D., \& {Goodman}, J. 2013, \bibinfo{title}{{emcee: The MCMC Hammer},} \pasp, 125, 306, \dodoi{10.1086/670067}

\bibitem[{A. {Frasca} {et~al.}(2009){Frasca}, {Covino}, {Spezzi}, {Alcal{\'a}}, {Marilli}, {F{\.z}r{\'e}sz}, \& {Gandolfi}}]{Frasca2009}
{Frasca}, A., {Covino}, E., {Spezzi}, L., {et~al.} 2009, \bibinfo{title}{{REM near-IR and optical photometric monitoring of pre-main sequence stars in Orion. Rotation periods and starspot parameters},} \aap, 508, 1313, \dodoi{10.1051/0004-6361/200913327}

\bibitem[{A. {Fukui} {et~al.}(2011){Fukui}, {Narita}, {Tristram}, {Sumi}, {Abe}, {Itow}, {Sullivan}, {Bond}, {Hirano}, {Tamura}, {Bennett}, {Furusawa}, {Hayashi}, {Hearnshaw}, {Hosaka}, {Kamiya}, {Kobara}, {Korpela}, {Kilmartin}, {Lin}, {Ling}, {Makita}, {Masuda}, {Matsubara}, {Miyake}, {Muraki}, {Nagaya}, {Nishimoto}, {Ohnishi}, {Omori}, {Perrott}, {Rattenbury}, {Saito}, {Skuljan}, {Suzuki}, {Sweatman}, \& {Wada}}]{Fukui2011}
{Fukui}, A., {Narita}, N., {Tristram}, P.~J., {et~al.} 2011, \bibinfo{title}{{Measurements of Transit Timing Variations for WASP-5b},} \pasj, 63, 287, \dodoi{10.1093/pasj/63.1.287}

\bibitem[{ {Gaia Collaboration} {et~al.}(2016){Gaia Collaboration}, {Prusti}, {de Bruijne}, {Brown}, {Vallenari}, {Babusiaux}, {Bailer-Jones}, {Bastian}, {Biermann}, {Evans}, {Eyer}, {Jansen}, {Jordi}, {Klioner}, {Lammers}, {Lindegren}, {Luri}, {Mignard}, {Milligan}, {Panem}, {Poinsignon}, {Pourbaix}, {Randich}, {Sarri}, {Sartoretti}, {Siddiqui}, {Soubiran}, {Valette}, {van Leeuwen}, {Walton}, {Aerts}, {Arenou}, {Cropper}, {Drimmel}, {H{\o}g}, {Katz}, {Lattanzi}, {O'Mullane}, {Grebel}, {Holland}, {Huc}, {Passot}, {Bramante}, {Cacciari}, {Casta{\~n}eda}, {Chaoul}, {Cheek}, {De Angeli}, {Fabricius}, {Guerra}, {Hern{\'a}ndez}, {Jean-Antoine-Piccolo}, {Masana}, {Messineo}, {Mowlavi}, {Nienartowicz}, {Ord{\'o}{\~n}ez-Blanco}, {Panuzzo}, {Portell}, {Richards}, {Riello}, {Seabroke}, {Tanga}, {Th{\'e}venin}, {Torra}, {Els}, {Gracia-Abril}, {Comoretto}, {Garcia-Reinaldos}, {Lock}, {Mercier}, {Altmann}, {Andrae}, {Astraatmadja}, {Bellas-Velidis}, {Benson}, {Berthier}, {Blomme}, {Busso}, {Carry}, {Cellino},
  {Clementini}, {Cowell}, {Creevey}, {Cuypers}, {Davidson}, {De Ridder}, {de Torres}, {Delchambre}, {Dell'Oro}, {Ducourant}, {Fr{\'e}mat}, {Garc{\'\i}a-Torres}, {Gosset}, {Halbwachs}, {Hambly}, {Harrison}, {Hauser}, {Hestroffer}, {Hodgkin}, {Huckle}, {Hutton}, {Jasniewicz}, {Jordan}, {Kontizas}, {Korn}, {Lanzafame}, {Manteiga}, {Moitinho}, {Muinonen}, {Osinde}, {Pancino}, {Pauwels}, {Petit}, {Recio-Blanco}, {Robin}, {Sarro}, {Siopis}, {Smith}, {Smith}, {Sozzetti}, {Thuillot}, {van Reeven}, {Viala}, {Abbas}, {Abreu Aramburu}, {Accart}, {Aguado}, {Allan}, {Allasia}, {Altavilla}, {{\'A}lvarez}, {Alves}, {Anderson}, {Andrei}, {Anglada Varela}, {Antiche}, {Antoja}, {Ant{\'o}n}, {Arcay}, {Atzei}, {Ayache}, {Bach}, {Baker}, {Balaguer-N{\'u}{\~n}ez}, {Barache}, {Barata}, {Barbier}, {Barblan}, {Baroni}, {Barrado y Navascu{\'e}s}, {Barros}, {Barstow}, {Becciani}, {Bellazzini}, {Bellei}, {Bello Garc{\'\i}a}, {Belokurov}, {Bendjoya}, {Berihuete}, {Bianchi}, {Bienaym{\'e}}, {Billebaud}, {Blagorodnova}, {Blanco-Cuaresma},
  {Boch}, {Bombrun}, {Borrachero}, {Bouquillon}, {Bourda}, {Bouy}, {Bragaglia}, {Breddels}, {Brouillet}, {Br{\"u}semeister}, {Bucciarelli}, {Budnik}, {Burgess}, {Burgon}, {Burlacu}, {Busonero}, {Buzzi}, {Caffau}, {Cambras}, {Campbell}, {Cancelliere}, {Cantat-Gaudin}, {Carlucci}, {Carrasco}, {Castellani}, {Charlot}, {Charnas}, {Charvet}, {Chassat}, {Chiavassa}, {Clotet}, {Cocozza}, {Collins}, {Collins}, \& {Costigan}}]{Gaia2016}
{Gaia Collaboration}, {Prusti}, T., {de Bruijne}, J.~H.~J., {et~al.} 2016, \bibinfo{title}{{The Gaia mission},} \aap, 595, A1, \dodoi{10.1051/0004-6361/201629272}

\bibitem[{ {Gaia Collaboration} {et~al.}(2023){Gaia Collaboration}, {Vallenari}, {Brown}, {Prusti}, {de Bruijne}, {Arenou}, {Babusiaux}, {Biermann}, {Creevey}, {Ducourant}, {Evans}, {Eyer}, {Guerra}, {Hutton}, {Jordi}, {Klioner}, {Lammers}, {Lindegren}, {Luri}, {Mignard}, {Panem}, {Pourbaix}, {Randich}, {Sartoretti}, {Soubiran}, {Tanga}, {Walton}, {Bailer-Jones}, {Bastian}, {Drimmel}, {Jansen}, {Katz}, {Lattanzi}, {van Leeuwen}, {Bakker}, {Cacciari}, {Casta{\~n}eda}, {De Angeli}, {Fabricius}, {Fouesneau}, {Fr{\'e}mat}, {Galluccio}, {Guerrier}, {Heiter}, {Masana}, {Messineo}, {Mowlavi}, {Nicolas}, {Nienartowicz}, {Pailler}, {Panuzzo}, {Riclet}, {Roux}, {Seabroke}, {Sordo}, {Th{\'e}venin}, {Gracia-Abril}, {Portell}, {Teyssier}, {Altmann}, {Andrae}, {Audard}, {Bellas-Velidis}, {Benson}, {Berthier}, {Blomme}, {Burgess}, {Busonero}, {Busso}, {C{\'a}novas}, {Carry}, {Cellino}, {Cheek}, {Clementini}, {Damerdji}, {Davidson}, {de Teodoro}, {Nu{\~n}ez Campos}, {Delchambre}, {Dell'Oro}, {Esquej},
  {Fern{\'a}ndez-Hern{\'a}ndez}, {Fraile}, {Garabato}, {Garc{\'\i}a-Lario}, {Gosset}, {Haigron}, {Halbwachs}, {Hambly}, {Harrison}, {Hern{\'a}ndez}, {Hestroffer}, {Hodgkin}, {Holl}, {Jan{\ss}en}, {Jevardat de Fombelle}, {Jordan}, {Krone-Martins}, {Lanzafame}, {L{\"o}ffler}, {Marchal}, {Marrese}, {Moitinho}, {Muinonen}, {Osborne}, {Pancino}, {Pauwels}, {Recio-Blanco}, {Reyl{\'e}}, {Riello}, {Rimoldini}, {Roegiers}, {Rybizki}, {Sarro}, {Siopis}, {Smith}, {Sozzetti}, {Utrilla}, {van Leeuwen}, {Abbas}, {{\'A}brah{\'a}m}, {Abreu Aramburu}, {Aerts}, {Aguado}, {Ajaj}, {Aldea-Montero}, {Altavilla}, {{\'A}lvarez}, {Alves}, {Anders}, {Anderson}, {Anglada Varela}, {Antoja}, {Baines}, {Baker}, {Balaguer-N{\'u}{\~n}ez}, {Balbinot}, {Balog}, {Barache}, {Barbato}, {Barros}, {Barstow}, {Bartolom{\'e}}, {Bassilana}, {Bauchet}, {Becciani}, {Bellazzini}, {Berihuete}, {Bernet}, {Bertone}, {Bianchi}, {Binnenfeld}, {Blanco-Cuaresma}, {Blazere}, {Boch}, {Bombrun}, {Bossini}, {Bouquillon}, {Bragaglia}, {Bramante}, {Breedt},
  {Bressan}, {Brouillet}, {Brugaletta}, {Bucciarelli}, {Burlacu}, {Butkevich}, {Buzzi}, {Caffau}, {Cancelliere}, {Cantat-Gaudin}, {Carballo}, {Carlucci}, {Carnerero}, {Carrasco}, {Casamiquela}, {Castellani}, {Castro-Ginard}, {Chaoul}, {Charlot}, {Chemin}, {Chiaramida}, {Chiavassa}, {Chornay}, {Comoretto}, {Contursi}, {Cooper}, {Cornez}, {Cowell}, {Crifo}, {Cropper}, {Crosta}, {Crowley}, {Dafonte}, {Dapergolas}, {David}, {David}, {de Laverny}, {De Luise}, \& {De March}}]{Gaia2023}
{Gaia Collaboration}, {Vallenari}, A., {Brown}, A.~G.~A., {et~al.} 2023, \bibinfo{title}{{Gaia Data Release 3. Summary of the content and survey properties},} \aap, 674, A1, \dodoi{10.1051/0004-6361/202243940}

\bibitem[{E. {Gaidos} {et~al.}(2020){Gaidos}, {Hirano}, {Wilson}, {France}, {Rockcliffe}, {Newton}, {Feiden}, {Krishnamurthy}, {Harakawa}, {Hodapp}, {Ishizuka}, {Jacobson}, {Konishi}, {Kotani}, {Kudo}, {Kurokawa}, {Kuzuhara}, {Nishikawa}, {Omiya}, {Serizawa}, {Tamura}, {Ueda}, \& {Vievard}}]{Gaidos2020}
{Gaidos}, E., {Hirano}, T., {Wilson}, D.~J., {et~al.} 2020, \bibinfo{title}{{Zodiacal exoplanets in time - XI. The orbit and radiation environment of the young M dwarf-hosted planet K2-25b},} \mnras, 498, L119, \dodoi{10.1093/mnrasl/slaa136}

\bibitem[{J.~P. {Gardner} {et~al.}(2006){Gardner}, {Mather}, {Clampin}, {Doyon}, {Greenhouse}, {Hammel}, {Hutchings}, {Jakobsen}, {Lilly}, {Long}, {Lunine}, {McCaughrean}, {Mountain}, {Nella}, {Rieke}, {Rieke}, {Rix}, {Smith}, {Sonneborn}, {Stiavelli}, {Stockman}, {Windhorst}, \& {Wright}}]{Gardner2006}
{Gardner}, J.~P., {Mather}, J.~C., {Clampin}, M., {et~al.} 2006, \bibinfo{title}{{The James Webb Space Telescope},} \ssr, 123, 485, \dodoi{10.1007/s11214-006-8315-7}

\bibitem[{A. {Ginsburg} {et~al.}(2019){Ginsburg}, {Sip{\H{o}}cz}, {Brasseur}, {Cowperthwaite}, {Craig}, {Deil}, {Guillochon}, {Guzman}, {Liedtke}, {Lian Lim}, {Lockhart}, {Mommert}, {Morris}, {Norman}, {Parikh}, {Persson}, {Robitaille}, {Segovia}, {Singer}, {Tollerud}, {de Val-Borro}, {Valtchanov}, {Woillez}, {Astroquery Collaboration}, \& {a subset of astropy Collaboration}}]{Astroquery2019}
{Ginsburg}, A., {Sip{\H{o}}cz}, B.~M., {Brasseur}, C.~E., {et~al.} 2019, \bibinfo{title}{{astroquery: An Astronomical Web-querying Package in Python},} \aj, 157, 98, \dodoi{10.3847/1538-3881/aafc33}

\bibitem[{C.~R. Harris {et~al.}(2020)Harris, Millman, van~der Walt, Gommers, Virtanen, Cournapeau, Wieser, Taylor, Berg, Smith, Kern, Picus, Hoyer, van Kerkwijk, Brett, Haldane, del R{'{\i}}o, Wiebe, Peterson, G{'{e}}rard-Marchant, Sheppard, Reddy, Weckesser, Abbasi, Gohlke, \& Oliphant}]{numpy2020}
Harris, C.~R., Millman, K.~J., van~der Walt, S.~J., {et~al.} 2020, \bibinfo{title}{Array programming with {NumPy},} Nature, 585, 357, \dodoi{10.1038/s41586-020-2649-2}

\bibitem[{K. {Herbst} {et~al.}(2021){Herbst}, {Papaioannou}, {Airapetian}, \& {Atri}}]{Herbst2021}
{Herbst}, K., {Papaioannou}, A., {Airapetian}, V.~S., \& {Atri}, D. 2021, \bibinfo{title}{{From Starspots to Stellar Coronal Mass Ejections{\textemdash}Revisiting Empirical Stellar Relations},} \apj, 907, 89, \dodoi{10.3847/1538-4357/abcc04}

\bibitem[{T. {Hirano} {et~al.}(2024){Hirano}, {Gaidos}, {Harakawa}, {Hodapp}, {Kotani}, {Kudo}, {Kurokawa}, {Kuzuhara}, {Mann}, {Nishikawa}, {Omiya}, {Serizawa}, {Tamura}, {Thao}, {Ueda}, \& {Vievard}}]{Hirano2024}
{Hirano}, T., {Gaidos}, E., {Harakawa}, H., {et~al.} 2024, \bibinfo{title}{{Transit spectroscopy of K2-33b with subaru/IRD: Spin-Orbit alignment and tentative atmospheric helium},} \mnras, 530, 3117, \dodoi{10.1093/mnras/stae998}

\bibitem[{J.~D. Hunter(2007)Hunter}]{matplotlib2007}
Hunter, J.~D. 2007, \bibinfo{title}{Matplotlib: A 2D graphics environment,} Computing in Science \& Engineering, 9, 90, \dodoi{10.1109/MCSE.2007.55}

\bibitem[{T.-O. Husser {et~al.}(2013)Husser, {Wende-von Berg}, Dreizler, Homeier, Reiners, Barman, \& Hauschildt}]{Husser2013}
Husser, T.-O., {Wende-von Berg}, S., Dreizler, S., {et~al.} 2013, \bibinfo{title}{{Astrophysics A new extensive library of PHOENIX stellar atmospheres},} A{\&}A, 553, A6, \dodoi{10.1051/0004-6361/201219058}

\bibitem[{K. {Ikuta} {et~al.}(2023){Ikuta}, {Namekata}, {Notsu}, {Maehara}, {Okamoto}, {Honda}, {Nogami}, \& {Shibata}}]{Ikuta2023}
{Ikuta}, K., {Namekata}, K., {Notsu}, Y., {et~al.} 2023, \bibinfo{title}{{Starspot Mapping with Adaptive Parallel Tempering. II. Application to TESS Data for M-dwarf Flare Stars AU Microscopii, YZ Canis Minoris, and EV Lacertae},} \apj, 948, 64, \dodoi{10.3847/1538-4357/acbd36}

\bibitem[{J.~M. {Jenkins} {et~al.}(2016){Jenkins}, {Twicken}, {McCauliff}, {Campbell}, {Sanderfer}, {Lung}, {Mansouri-Samani}, {Girouard}, {Tenenbaum}, {Klaus}, {Smith}, {Caldwell}, {Chacon}, {Henze}, {Heiges}, {Latham}, {Morgan}, {Swade}, {Rinehart}, \& {Vanderspek}}]{Jenkins2016}
{Jenkins}, J.~M., {Twicken}, J.~D., {McCauliff}, S., {et~al.} 2016, in Society of Photo-Optical Instrumentation Engineers (SPIE) Conference Series, Vol. 9913, Software and Cyberinfrastructure for Astronomy IV, ed. G.~{Chiozzi} \& J.~C. {Guzman}, 99133E, \dodoi{10.1117/12.2233418}

\bibitem[{D.~M. {Kipping}(2012){Kipping}}]{Kipping2012}
{Kipping}, D.~M. 2012, \bibinfo{title}{{An analytic model for rotational modulations in the photometry of spotted stars},} \mnras, 427, 2487, \dodoi{10.1111/j.1365-2966.2012.22124.x}

\bibitem[{D.~M. {Kipping}(2013){Kipping}}]{Kipping2013}
{Kipping}, D.~M. 2013, \bibinfo{title}{{Efficient, uninformative sampling of limb darkening coefficients for two-parameter laws},} \mnras, 435, 2152, \dodoi{10.1093/mnras/stt1435}

\bibitem[{T. Kluyver {et~al.}(2016)Kluyver, Ragan-Kelley, P{\'e}rez, Granger, Bussonnier, Frederic, Kelley, Hamrick, Grout, Corlay, Ivanov, Avila, Abdalla, \& Willing}]{jupyter2016}
Kluyver, T., Ragan-Kelley, B., P{\'e}rez, F., {et~al.} 2016, in Positioning and Power in Academic Publishing: Players, Agents and Agendas, ed. F.~Loizides \& B.~Schmidt, IOS Press, 87 -- 90

\bibitem[{O. {Kochukhov}(2021){Kochukhov}}]{Kochukhov2021}
{Kochukhov}, O. 2021, \bibinfo{title}{{Magnetic fields of M dwarfs},} \aapr, 29, 1, \dodoi{10.1007/s00159-020-00130-3}

\bibitem[{L. {Kreidberg}(2015){Kreidberg}}]{Kreidberg2015}
{Kreidberg}, L. 2015, \bibinfo{title}{{batman: BAsic Transit Model cAlculatioN in Python},} \pasp, 127, 1161, \dodoi{10.1086/683602}

\bibitem[{J.~E. {Libby-Roberts} {et~al.}(2023){Libby-Roberts}, {Schutte}, {Hebb}, {Kanodia}, {Ca{\~n}as}, {Stef{\'a}nsson}, {Lin}, {Mahadevan}, {Parts}, {Powers}, {Wisniewski}, {Bender}, {Cochran}, {Diddams}, {Everett}, {Gupta}, {Halverson}, {Kobulnicky}, {Kowalski}, {Larsen}, {Monson}, {Ninan}, {Parker}, {Ramsey}, {Robertson}, {Schwab}, {Swaby}, \& {Terrien}}]{Libby-Roberts2023}
{Libby-Roberts}, J.~E., {Schutte}, M., {Hebb}, L., {et~al.} 2023, \bibinfo{title}{{An In-depth Look at TOI-3884b: A Super-Neptune Transiting an M4Dwarf with Persistent Starspot Crossings},} \aj, 165, 249, \dodoi{10.3847/1538-3881/accc2f}

\bibitem[{ {Lightkurve Collaboration} {et~al.}(2018){Lightkurve Collaboration}, {Cardoso}, {Hedges}, {Gully-Santiago}, {Saunders}, {Cody}, {Barclay}, {Hall}, {Sagear}, {Turtelboom}, {Zhang}, {Tzanidakis}, {Mighell}, {Coughlin}, {Bell}, {Berta-Thompson}, {Williams}, {Dotson}, \& {Barentsen}}]{lk2018}
{Lightkurve Collaboration}, {Cardoso}, J.~V.~d.~M., {Hedges}, C., {et~al.} 2018, \bibinfo{title}{{Lightkurve: Kepler and TESS time series analysis in Python},}, Astrophysics Source Code Library \doeprint{1812.013}

\bibitem[{E.~M. {Louden} \& S.~C. {Millholland}(2024){Louden} \& {Millholland}}]{Louden2024}
{Louden}, E.~M., \& {Millholland}, S.~C. 2024, \bibinfo{title}{{Polar Neptunes Are Stable to Tides},} \apj, 974, 304, \dodoi{10.3847/1538-4357/ad74ff}

\bibitem[{S. {Mahadevan} {et~al.}(2012){Mahadevan}, {Ramsey}, {Bender}, {Terrien}, {Wright}, {Halverson}, {Hearty}, {Nelson}, {Burton}, {Redman}, {Osterman}, {Diddams}, {Kasting}, {Endl}, \& {Deshpande}}]{Mahadevan2012}
{Mahadevan}, S., {Ramsey}, L., {Bender}, C., {et~al.} 2012, in Society of Photo-Optical Instrumentation Engineers (SPIE) Conference Series, Vol. 8446, Ground-based and Airborne Instrumentation for Astronomy IV, ed. I.~S. {McLean}, S.~K. {Ramsay}, \& H.~{Takami}, 84461S, \dodoi{10.1117/12.926102}

\bibitem[{N.~I. {Mahmud} {et~al.}(2011){Mahmud}, {Crockett}, {Johns-Krull}, {Prato}, {Hartigan}, {Jaffe}, \& {Beichman}}]{Mahmud2011}
{Mahmud}, N.~I., {Crockett}, C.~J., {Johns-Krull}, C.~M., {et~al.} 2011, \bibinfo{title}{{Starspot-induced Optical and Infrared Radial Velocity Variability in T Tauri Star Hubble I 4},} \apj, 736, 123, \dodoi{10.1088/0004-637X/736/2/123}

\bibitem[{F.~J. {Masci} {et~al.}(2019){Masci}, {Laher}, {Rusholme}, {Shupe}, {Groom}, {Surace}, {Jackson}, {Monkewitz}, {Beck}, {Flynn}, {Terek}, {Landry}, {Hacopians}, {Desai}, {Howell}, {Brooke}, {Imel}, {Wachter}, {Ye}, {Lin}, {Cenko}, {Cunningham}, {Rebbapragada}, {Bue}, {Miller}, {Mahabal}, {Bellm}, {Patterson}, {Juri{\'c}}, {Golkhou}, {Ofek}, {Walters}, {Graham}, {Kasliwal}, {Dekany}, {Kupfer}, {Burdge}, {Cannella}, {Barlow}, {Van Sistine}, {Giomi}, {Fremling}, {Blagorodnova}, {Levitan}, {Riddle}, {Smith}, {Helou}, {Prince}, \& {Kulkarni}}]{ztf2}
{Masci}, F.~J., {Laher}, R.~R., {Rusholme}, B., {et~al.} 2019, \bibinfo{title}{{The Zwicky Transient Facility: Data Processing, Products, and Archive},} \pasp, 131, 018003, \dodoi{10.1088/1538-3873/aae8ac}

\bibitem[{C. {McCully} {et~al.}(2018){McCully}, {Volgenau}, {Harbeck}, {Lister}, {Saunders}, {Turner}, {Siiverd}, \& {Bowman}}]{McCully2018}
{McCully}, C., {Volgenau}, N.~H., {Harbeck}, D.-R., {et~al.} 2018, in Society of Photo-Optical Instrumentation Engineers (SPIE) Conference Series, Vol. 10707, Software and Cyberinfrastructure for Astronomy V, ed. J.~C. {Guzman} \& J.~{Ibsen}, 107070K, \dodoi{10.1117/12.2314340}

\bibitem[{K. {Miyakawa} {et~al.}(2021){Miyakawa}, {Hirano}, {Fukui}, {Mann}, {Gaidos}, \& {Sato}}]{Miyakawa2021}
{Miyakawa}, K., {Hirano}, T., {Fukui}, A., {et~al.} 2021, \bibinfo{title}{{Wavelength Dependence of Activity-induced Photometric Variations for Young Cool Stars in Hyades},} \aj, 162, 104, \dodoi{10.3847/1538-3881/ac111d}

\bibitem[{M. {Mori} {et~al.}(2024){Mori}, {Ikuta}, {Fukui}, {Narita}, {de Leon}, {Livingston}, {Ikoma}, {Kawai}, {Kawauchi}, {Murgas}, {Palle}, {Parviainen}, {Fern{\'a}ndez Rodr{\'\i}guez}, {Terada}, {Watanabe}, \& {Tamura}}]{Mori2024}
{Mori}, M., {Ikuta}, K., {Fukui}, A., {et~al.} 2024, \bibinfo{title}{{Characterization of starspots on a young M-dwarf K2-25: multiband observations of stellar photometric variability and planetary transits},} \mnras, 530, 167, \dodoi{10.1093/mnras/stae841}

\bibitem[{B. {Morris}(2020){Morris}}]{Morris2020}
{Morris}, B. 2020, \bibinfo{title}{{fleck: Fast approximate light curves for starspot rotational modulation},} The Journal of Open Source Software, 5, 2103, \dodoi{10.21105/joss.02103}

\bibitem[{C. {Moutou} {et~al.}(2017){Moutou}, {H{\'e}brard}, {Morin}, {Malo}, {Fouqu{\'e}}, {Torres-Rivas}, {Martioli}, {Delfosse}, {Artigau}, \& {Doyon}}]{Moutou2017}
{Moutou}, C., {H{\'e}brard}, E.~M., {Morin}, J., {et~al.} 2017, \bibinfo{title}{{SPIRou input catalogue: activity, rotation and magnetic field of cool dwarfs},} \mnras, 472, 4563, \dodoi{10.1093/mnras/stx2306}

\bibitem[{N. {Narita} {et~al.}(2020){Narita}, {Fukui}, {Yamamuro}, {Harbeck}, {Bowman}, {Elphick}, {Nation}, {Armstrong}, {Han}, {Abe}, {Ikoma}, {Isogai}, {Kawauchi}, {Kurita}, {Kusakabe}, {de Leon}, {Livingston}, {Mori}, {Nishiumi}, {Tamura}, {Watanabe}, {Volgenau}, {Heinrich-Josties}, {Foale}, {Daily}, {McCully}, {Kirby}, {Smith}, {Haworth}, {Conway}, {Storrie-Lombardi}, {Rosing}, {Chatelain}, {Bachelet}, {Johnson}, \& {Rabus}}]{Narita2020}
{Narita}, N., {Fukui}, A., {Yamamuro}, T., {et~al.} 2020, in Society of Photo-Optical Instrumentation Engineers (SPIE) Conference Series, Vol. 11447, Ground-based and Airborne Instrumentation for Astronomy VIII, ed. C.~J. {Evans}, J.~J. {Bryant}, \& K.~{Motohara}, 114475K, \dodoi{10.1117/12.2559947}

\bibitem[{H. Parviainen \& S. Aigrain(2015)Parviainen \& Aigrain}]{Parviainen2015}
Parviainen, H., \& Aigrain, S. 2015, \bibinfo{title}{{ldtk: Limb Darkening Toolkit},} MNRAS, 453, 3821, \dodoi{10.1093/mnras/stv1857}

\bibitem[{F. {Pepe} {et~al.}(2021){Pepe}, {Cristiani}, {Rebolo}, {Santos}, {Dekker}, {Cabral}, {Di Marcantonio}, {Figueira}, {Lo Curto}, {Lovis}, {Mayor}, {M{\'e}gevand}, {Molaro}, {Riva}, {Zapatero Osorio}, {Amate}, {Manescau}, {Pasquini}, {Zerbi}, {Adibekyan}, {Abreu}, {Affolter}, {Alibert}, {Aliverti}, {Allart}, {Allende Prieto}, {{\'A}lvarez}, {Alves}, {Avila}, {Baldini}, {Bandy}, {Barros}, {Benz}, {Bianco}, {Borsa}, {Bourrier}, {Bouchy}, {Broeg}, {Calderone}, {Cirami}, {Coelho}, {Conconi}, {Coretti}, {Cumani}, {Cupani}, {D'Odorico}, {Damasso}, {Deiries}, {Delabre}, {Demangeon}, {Dumusque}, {Ehrenreich}, {Faria}, {Fragoso}, {Genolet}, {Genoni}, {G{\'e}nova Santos}, {Gonz{\'a}lez Hern{\'a}ndez}, {Hughes}, {Iwert}, {Kerber}, {Knudstrup}, {Landoni}, {Lavie}, {Lillo-Box}, {Lizon}, {Maire}, {Martins}, {Mehner}, {Micela}, {Modigliani}, {Monteiro}, {Monteiro}, {Moschetti}, {Murphy}, {Nunes}, {Oggioni}, {Oliveira}, {Oshagh}, {Pall{\'e}}, {Pariani}, {Poretti}, {Rasilla}, {Rebord{\~a}o}, {Redaelli}, {Santana
  Tschudi}, {Santin}, {Santos}, {S{\'e}gransan}, {Schmidt}, {Segovia}, {Sosnowska}, {Sozzetti}, {Sousa}, {Span{\`o}}, {Su{\'a}rez Mascare{\~n}o}, {Tabernero}, {Tenegi}, {Udry}, \& {Zanutta}}]{Pepe2021}
{Pepe}, F., {Cristiani}, S., {Rebolo}, R., {et~al.} 2021, \bibinfo{title}{{ESPRESSO at VLT. On-sky performance and first results},} \aap, 645, A96, \dodoi{10.1051/0004-6361/202038306}

\bibitem[{C. {Petrovich} {et~al.}(2020){Petrovich}, {Mu{\~n}oz}, {Kratter}, \& {Malhotra}}]{Petrovich2020}
{Petrovich}, C., {Mu{\~n}oz}, D.~J., {Kratter}, K.~M., \& {Malhotra}, R. 2020, \bibinfo{title}{{A Disk-driven Resonance as the Origin of High Inclinations of Close-in Planets},} \apjl, 902, L5, \dodoi{10.3847/2041-8213/abb952}

\bibitem[{B.~V. {Rackham} {et~al.}(2018){Rackham}, {Apai}, \& {Giampapa}}]{Rackham2018}
{Rackham}, B.~V., {Apai}, D., \& {Giampapa}, M.~S. 2018, \bibinfo{title}{{The Transit Light Source Effect: False Spectral Features and Incorrect Densities for M-dwarf Transiting Planets},} \apj, 853, 122, \dodoi{10.3847/1538-4357/aaa08c}

\bibitem[{B.~V. {Rackham} {et~al.}(2023){Rackham}, {Espinoza}, {Berdyugina}, {Korhonen}, {MacDonald}, {Montet}, {Morris}, {Oshagh}, {Shapiro}, {Unruh}, {Quintana}, {Zellem}, {Apai}, {Barclay}, {Barstow}, {Bruno}, {Carone}, {Casewell}, {Cegla}, {Criscuoli}, {Fischer}, {Fournier}, {Giampapa}, {Giles}, {Iyer}, {Kopp}, {Kostogryz}, {Krivova}, {Mallonn}, {McGruder}, {Molaverdikhani}, {Newton}, {Panja}, {Peacock}, {Reardon}, {Roettenbacher}, {Scandariato}, {Solanki}, {Stassun}, {Steiner}, {Stevenson}, {Tregloan-Reed}, {Valio}, {Wedemeyer}, {Welbanks}, {Yu}, {Alam}, {Davenport}, {Deming}, {Dong}, {Ducrot}, {Fisher}, {Gilbert}, {Kostov}, {L{\'o}pez-Morales}, {Line}, {Mo{\v{c}}nik}, {Mullally}, {Paudel}, {Ribas}, \& {Valenti}}]{Rackham2023}
{Rackham}, B.~V., {Espinoza}, N., {Berdyugina}, S.~V., {et~al.} 2023, \bibinfo{title}{{The effect of stellar contamination on low-resolution transmission spectroscopy: needs identified by NASA's Exoplanet Exploration Program Study Analysis Group 21},} RAS Techniques and Instruments, 2, 148, \dodoi{10.1093/rasti/rzad009}

\bibitem[{G.~R. {Ricker} {et~al.}(2015){Ricker}, {Winn}, {Vanderspek}, {Latham}, {Bakos}, {Bean}, {Berta-Thompson}, {Brown}, {Buchhave}, {Butler}, {Butler}, {Chaplin}, {Charbonneau}, {Christensen-Dalsgaard}, {Clampin}, {Deming}, {Doty}, {De Lee}, {Dressing}, {Dunham}, {Endl}, {Fressin}, {Ge}, {Henning}, {Holman}, {Howard}, {Ida}, {Jenkins}, {Jernigan}, {Johnson}, {Kaltenegger}, {Kawai}, {Kjeldsen}, {Laughlin}, {Levine}, {Lin}, {Lissauer}, {MacQueen}, {Marcy}, {McCullough}, {Morton}, {Narita}, {Paegert}, {Palle}, {Pepe}, {Pepper}, {Quirrenbach}, {Rinehart}, {Sasselov}, {Sato}, {Seager}, {Sozzetti}, {Stassun}, {Sullivan}, {Szentgyorgyi}, {Torres}, {Udry}, \& {Villasenor}}]{Ricker2015}
{Ricker}, G.~R., {Winn}, J.~N., {Vanderspek}, R., {et~al.} 2015, \bibinfo{title}{{Transiting Exoplanet Survey Satellite (TESS)},} Journal of Astronomical Telescopes, Instruments, and Systems, 1, 014003, \dodoi{10.1117/1.JATIS.1.1.014003}

\bibitem[{R. {Sanchis-Ojeda} \& J.~N. {Winn}(2011){Sanchis-Ojeda} \& {Winn}}]{Sanchis-Ojeda2011}
{Sanchis-Ojeda}, R., \& {Winn}, J.~N. 2011, \bibinfo{title}{{Starspots, Spin-Orbit Misalignment, and Active Latitudes in the HAT-P-11 Exoplanetary System},} \apj, 743, 61, \dodoi{10.1088/0004-637X/743/1/61}

\bibitem[{M.~C. {Schutte} {et~al.}(2023){Schutte}, {Hebb}, {Wisniewski}, {Ca{\~n}as}, {Libby-Roberts}, {Lin}, {Robertson}, \& {Stef{\'a}nsson}}]{Schutte2023}
{Schutte}, M.~C., {Hebb}, L., {Wisniewski}, J.~P., {et~al.} 2023, \bibinfo{title}{{Measuring the Temperature of Starspots from Multi-filter Photometry},} \aj, 166, 92, \dodoi{10.3847/1538-3881/ace59c}

\bibitem[{A. {Shporer} {et~al.}(2024){Shporer}, {Collins}, {Watkins}, {Barkaoui}, {Conti}, {Jensen}, {Johnson}, {Massey}, {Relles}, {Schwarz}, \& {Stockdale}}]{Shporer2024}
{Shporer}, A., {Collins}, K., {Watkins}, C., {et~al.} 2024, in TESS Science Conference III, 27, \dodoi{10.5281/zenodo.12803734}

\bibitem[{J. {Southworth}(2011){Southworth}}]{Southworth2011}
{Southworth}, J. 2011, \bibinfo{title}{{Homogeneous studies of transiting extrasolar planets - IV. Thirty systems with space-based light curves},} \mnras, 417, 2166, \dodoi{10.1111/j.1365-2966.2011.19399.x}

\bibitem[{G. {Stefansson} {et~al.}(2020){Stefansson}, {Mahadevan}, {Maney}, {Ninan}, {Robertson}, {Rajagopal}, {Haase}, {Allen}, {Ford}, {Winn}, {Wolfgang}, {Dawson}, {Wisniewski}, {Bender}, {Ca{\~n}as}, {Cochran}, {Diddams}, {Fredrick}, {Halverson}, {Hearty}, {Hebb}, {Kanodia}, {Levi}, {Metcalf}, {Monson}, {Ramsey}, {Roy}, {Schwab}, {Terrien}, \& {Wright}}]{Stefansson2020}
{Stefansson}, G., {Mahadevan}, S., {Maney}, M., {et~al.} 2020, \bibinfo{title}{{The Habitable Zone Planet Finder Reveals a High Mass and Low Obliquity for the Young Neptune K2-25b},} \aj, 160, 192, \dodoi{10.3847/1538-3881/abb13a}

\bibitem[{G. {Stef{\`a}nsson} {et~al.}(2022){Stef{\`a}nsson}, {Mahadevan}, {Petrovich}, {Winn}, {Kanodia}, {Millholland}, {Maney}, {Ca{\~n}as}, {Wisniewski}, {Robertson}, {Ninan}, {Ford}, {Bender}, {Blake}, {Cegla}, {Cochran}, {Diddams}, {Dong}, {Endl}, {Fredrick}, {Halverson}, {Hearty}, {Hebb}, {Hirano}, {Lin}, {Logsdon}, {Lubar}, {McElwain}, {Metcalf}, {Monson}, {Rajagopal}, {Ramsey}, {Roy}, {Schwab}, {Schweiker}, {Terrien}, \& {Wright}}]{Stefansson2022}
{Stef{\`a}nsson}, G., {Mahadevan}, S., {Petrovich}, C., {et~al.} 2022, \bibinfo{title}{{The Warm Neptune GJ 3470b Has a Polar Orbit},} \apjl, 931, L15, \dodoi{10.3847/2041-8213/ac6e3c}

\bibitem[{K.~G. {Strassmeier}(2009){Strassmeier}}]{Strassmeier2009}
{Strassmeier}, K.~G. 2009, \bibinfo{title}{{Starspots},} \aapr, 17, 251, \dodoi{10.1007/s00159-009-0020-6}

\bibitem[{C.~J.~F. ter Braak \& J.~A. Vrugt(2008)ter Braak \& Vrugt}]{Braak2008}
ter Braak, C. J.~F., \& Vrugt, J.~A. 2008, \bibinfo{title}{Differential Evolution Markov Chain with snooker updater and fewer chains,} Statistics and Computing, 18, 435–446, \dodoi{10.1007/s11222-008-9104-9}

\bibitem[{P.~C. {Thao} {et~al.}(2023){Thao}, {Mann}, {Gao}, {Owens}, {Vanderburg}, {Newton}, {Tang}, {Fields}, {David}, {Irwin}, {Husser}, {Charbonneau}, \& {Ballard}}]{Thao2023}
{Thao}, P.~C., {Mann}, A.~W., {Gao}, P., {et~al.} 2023, \bibinfo{title}{{Hazy with a Chance of Star Spots: Constraining the Atmosphere of Young Planet K2-33b},} \aj, 165, 23, \dodoi{10.3847/1538-3881/aca07a}

\bibitem[{ {The pandas development Team}(2024){The pandas development Team}}]{pandas2022}
{The pandas development Team}. 2024, \bibinfo{title}{{pandas-dev/pandas: Pandas},}, v2.2.3 Zenodo, \dodoi{10.5281/zenodo.3509134}

\bibitem[{A. {Thompson} {et~al.}(2024){Thompson}, {Biagini}, {Cracchiolo}, {Petralia}, {Changeat}, {Saba}, {Morello}, {Morvan}, {Micela}, \& {Tinetti}}]{Thompson2024}
{Thompson}, A., {Biagini}, A., {Cracchiolo}, G., {et~al.} 2024, \bibinfo{title}{{Correcting Exoplanet Transmission Spectra for Stellar Activity with an Optimized Retrieval Framework},} \apj, 960, 107, \dodoi{10.3847/1538-4357/ad0369}

\bibitem[{P. {Virtanen} {et~al.}(2020){Virtanen}, {Gommers}, {Oliphant}, {Haberland}, {Reddy}, {Cournapeau}, {Burovski}, {Peterson}, {Weckesser}, {Bright}, {van der Walt}, {Brett}, {Wilson}, {Millman}, {Mayorov}, {Nelson}, {Jones}, {Kern}, {Larson}, {Carey}, {Polat}, {Feng}, {Moore}, {VanderPlas}, {Laxalde}, {Perktold}, {Cimrman}, {Henriksen}, {Quintero}, {Harris}, {Archibald}, {Ribeiro}, {Pedregosa}, {van Mulbregt}, \& {SciPy 1. 0 Contributors}}]{scipy2020}
{Virtanen}, P., {Gommers}, R., {Oliphant}, T.~E., {et~al.} 2020, \bibinfo{title}{{SciPy 1.0: fundamental algorithms for scientific computing in Python},} Nature Methods, 17, 261, \dodoi{10.1038/s41592-019-0686-2}

\bibitem[{W.~C. {Waalkes} {et~al.}(2024){Waalkes}, {Berta-Thompson}, {Newton}, {Mann}, {Gao}, {Wakeford}, {Alderson}, \& {Plavchan}}]{Waalkes2024}
{Waalkes}, W.~C., {Berta-Thompson}, Z.~K., {Newton}, E.~R., {et~al.} 2024, \bibinfo{title}{{Quantifying the Transit Light Source Effect: Measurements of Spot Temperature and Coverage on the Photosphere of AU Microscopii with High-resolution Spectroscopy and Multicolor Photometry},} \apj, 962, 97, \dodoi{10.3847/1538-4357/ad0bed}

\bibitem[{N.~J. {Wright} {et~al.}(2018){Wright}, {Newton}, {Williams}, {Drake}, \& {Yadav}}]{Wright2018}
{Wright}, N.~J., {Newton}, E.~R., {Williams}, P. K.~G., {Drake}, J.~J., \& {Yadav}, R.~K. 2018, \bibinfo{title}{{The stellar rotation-activity relationship in fully convective M dwarfs},} \mnras, 479, 2351, \dodoi{10.1093/mnras/sty1670}

\bibitem[{R.~K. {Yadav} {et~al.}(2015){Yadav}, {Gastine}, {Christensen}, \& {Reiners}}]{Yadav2015}
{Yadav}, R.~K., {Gastine}, T., {Christensen}, U.~R., \& {Reiners}, A. 2015, \bibinfo{title}{{Formation of starspots in self-consistent global dynamo models: Polar spots on cool stars},} \aap, 573, A68, \dodoi{10.1051/0004-6361/201424589}

\bibitem[{M. {Zechmeister} \& M. {K{\"u}rster}(2009){Zechmeister} \& {K{\"u}rster}}]{GLS}
{Zechmeister}, M., \& {K{\"u}rster}, M. 2009, \bibinfo{title}{{The generalised Lomb-Scargle periodogram. A new formalism for the floating-mean and Keplerian periodograms},} \aap, 496, 577, \dodoi{10.1051/0004-6361:200811296}

\bibitem[{ {ZTF Team}(2025){ZTF Team}}]{https://doi.org/10.26131/irsa598}
{ZTF Team}. 2025, \bibinfo{title}{ZTF Lightcurves,} IPAC, \dodoi{10.26131/IRSA598}

\end{thebibliography}
\bibliographystyle{aasjournal}

\end{document}